\documentclass[fleqn,usenatbib]{mnras}

\usepackage{newtxtext,newtxmath}
\usepackage{stfloats}
\usepackage[flushleft]{threeparttable}

\usepackage[T1]{fontenc}
\usepackage{setspace}

\usepackage{graphicx}	
\usepackage{amsmath}	

\usepackage{amssymb}	
\usepackage[justification=centering]{caption}
\usepackage{float}

\title[Periodic X-ray Sources in the Galactic Bulge]{Periodic X-ray Sources in the Galactic Bulge: Application of the Gregory-Loredo Algorithm}

\author[Bao \& Li]{
Tong Bao$^{1,2}$\thanks{E-mail: baotong@smail.nju.edu.cn}
Zhiyuan Li$^{1,2}$\thanks{E-mail: lizy@nju.edu.cn}
\\
$^{1}$School of Astronomy and Space Science, Nanjing University, Nanjing 210046, China\\
$^{2}$Key Laboratory of Modern Astronomy and Astrophysics (Nanjing University), Ministry of Education, Nanjing 210046, China
}

\date{Accepted 2020 August 24. Received 2020 August 16; in original form 2020 July 09}

\pubyear{2020}

\begin{document}
\maketitle
\begin{abstract}
We present a systematic study of periodic X-ray sources in the Limiting Window (LW), a $\sim$70 arcmin$^2$ field representative of the inner Galactic bulge and the target of $\sim$1 Ms {\it Chandra} observations.  
Using the Gregory-Loredo algorithm, which applies Bayes's theorem to the phase-folded light curve  and is well-suited for irregularly sampled X-ray data, 
we detect 25 periodic signals in 23 discrete sources, among which 15 signals are new discoveries and two sources show dual periods. 
The vast majority of the 23 periodic sources are classified as magnetic cataclysmic variables (CVs), based on their period range, X-ray luminosities, spectral properties, and phase-folded light curves that are characteristic of spin modulation.
Meanwhile, there is a paucity of non-magnetic CVs seen as periodic sources, which can be understood as due to a low detection efficiency for eclipsing sources. 
Under reasonable assumptions about the geometry of magnetic CVs and a large set of simulated X-ray light curves, we estimate the fraction of magnetic CVs in the inner Galactic bulge to be $\lesssim$23\%, which is similar to that in the solar neighborhood.
There is an apparent lack of long-period ($\gtrsim$3.3 hours) CVs in the LW, when contrasted with the range of known CVs in the solar neighborhood.
We suggest that this might be an age effect, in the sense that CVs in the inner bulge are more evolved systems and have substantially shrunk their orbits. 
\end{abstract}


\section{Introduction} \label{sec:intro}
Mass-transferring, close binaries, which involve a black hole (BH), a neutron star (NS) or a white dwarf (WD) accreting from a companion star, are among the first objects discovered in the X-ray sky and now understood to be ubiquitous in the local universe, in particular our own Galaxy \citep{2006ARA&A..44..323F}. 
As such, X-ray binaries can serve as a useful probe of their parent stellar populations.

Due to their relative dimness in the quiescent state,
well-studied examples of cataclysmic variables (CVs), close binaries consisting of a WD accretor and a main-sequence or sub-giant donor, mainly reside in the solar neighborhood. 
Popular catalogs of CVs now include more than one thousand sources, most with well determined physical properties of the binary system (\citealp{2001PASP..113..764D,2003A&A...404..301R}).
Most known CVs have an orbital period between 1--12 hours.
The evolution of CVs is driven by angular momentum loss (AML) that leads to a gradual shrinking of the binary orbit.
The dominant AML mechanism in systems of long orbital periods ($P_{\rm orb}$ $\gtrsim$ 3 hours) is magnetic braking, whereas in short-period CVs ($P_{\rm orb}$ $\lesssim$ 2 hour) the AML is driven by gravitational radiation. There exists a ``period gap'' between about 2--3 hours, within which mass transfer is largely suppressed and few CVs are found. CVs also exhibit a minimum orbital period of $\sim$80 minutes, which is the result of the donor becoming a degenerate star due to cumulative mass loss. 
 
CVs are dubbed magnetic or non-magnetic, according to whether the WD has a strong surface magnetic field. Magnetic CVs can be further divided into polars and intermediate polars (IPs), depending on the level of synchronization between the orbital period ($P_{\rm orb}$) and spin period ($P_{\rm spin}$) of the WD. Polars have a near-perfect synchronization, $P_{\rm spin}/P_{\rm orb}\simeq 1$, a result of magnetic coupling owing to a strong magnetic field ($\gtrsim$ 10 MG on the WD surface).
IPs, on the other hand, have weaker surface magnetic fields (typically 1--10 MG) and are thus less synchronized ($P_{\rm spin}/P_{\rm orb}\simeq 0.01-1$).
Empirically, the fraction of magnetic CVs (polars plus IPs) is $\sim$20\% among all known CVs in the solar neighborhood (\citealt{2003A&A...404..301R}), while the intrinsic fraction might be somewhat lower ($\sim$16\%; \citealp{2013MNRAS.432..570P}).
However, the origin of the strong magnetic field in CVs remains unclear. 

While historically the CV phenomenon roots on the optical band, X-ray observations have been providing complementary and crucial knowledge about CVs, since the first CV detection in the X-ray band (SS Cyg, \citealp{1974ApJ...187L...5R,1978ApJ...224..953S}). 
CVs exhibit persistent X-ray luminosities ranging between $10^{29}-10^{34}\rm~erg~s^{-1}$, with short-term and long-term variability. 
X-rays from a CV are almost exclusively produced near the WD surface (\citealp{2001cvs..book.....H, 2017PASP..129f2001M}). 
In the case of non-magnetic CVs (dominated by dwarf novae [DNe]), accretion proceeds through a disk, releasing about half of the gravitational energy in heating the disk and the other half in a boundary layer between the WD surface and the inner edge of the disk, where the accreted material drastically decelerates and forms a shock. It is the post-shock plasma that produces the optically-thin thermal X-ray emission.
In the case of magnetic CVs, the accreted material is constrained by magnetic field lines, landing near the magnetic poles in the form of an accretion stream or curtain, creating an accretion shock, and producing copious X-rays in the post-shock plasma. 
This particular geometry of magnetic CVs often leads to spin-orbital modulations on the X-ray light curve, which can be a useful diagnostic of magnetic CVs. 

The advent of {\it Chandra} observations have opened up the possibility of detecting even the least luminous CVs in the inner Galactic regions, which is inaccessible to optical observations and suffers from strong crowding effect. 
In particular, it has been suggested that thousands of discrete X-ray sources detected in the innermost degree of the Galactic center are predominantly CVs \citep{2002Natur.415..148W,2003ApJ...589..225M,2006ApJS..165..173M,2009ApJS..181..110M,2018ApJS..235...26Z}. 
Toward a so-called {\it Limiting Window} (LW) of relatively low line-of-sight extinction, which samples the inner Galactic bulge at a projected distance of $\sim1{\fdg}4$ from the Galactic center, deep {\it Chandra} observations also resolved several hundreds of discrete sources  \citep{2009Natur.458.1142R}, the majority of which are thought to be CVs and coronally active binaries (ABs). 


The physical nature of the CVs in the Galactic center and LW, however, remains elusive. 
It has been suggested that the majority of detected CVs in the Galactic center are IPs, based on several arguments: (i) Their typical X-ray spectra showing a hard continuum and significant Fe lines are characteristic of IPs \citep{2004ApJ...613.1179M,2009ApJS..181..110M}; (ii) The extended 20–40 keV emission from the Galactic center, recently discovered by NuSTAR \citep{2015Natur.520..646P}, is consistent with originating from thousands of IPs having an average WD mass of $0.9\rm~M_\odot$ 
\citep{2016ApJ...826..160H};
(iii) Eight sources in the Galactic center are found to show periodic flux modulations, with periods consistent with magnetic CVs \citep{2003ApJ...599..465M}. 
Similarly, ten periodic sources were identified in the LW and suggested to be 
polars due to their period distribution \citep{2012ApJ...746..165H}. 
The prevalence of magnetic CVs in the Galactic center and the inner Galactic bulge appears at odds with the low fraction ($\sim$20\%) of magnetic CVs in the solar neighborhood. This may indicate that either the inner Galaxy hosts a significantly different population of CVs, or a large number of non-magnetic CVs thereof still awaits discovery.
Indeed, based on the measured Fe line ratios in the cumulative spectrum of resolved X-ray sources in the Galactic center and the LW, \citet{2018ApJS..235...26Z} found evidence that at least a fraction of these sources are likely DNe (see also \citealp{2016ApJ...818..136X}).


In this work, we revisit the deep {\it Chandra} observations of the LW, employing a novel technique, known as the Gregory-Loredo period searching algorithm \citep{1992ApJ...398..146G}, to detect periodic sources. We will show that this enables us to find more periodic X-ray sources than previous work using the same dataset \citep{2012ApJ...746..165H}, and that from the classification of these sources the fraction of magnetic CVs in the LW can be reasonably estimated. 
The rest of this paper is structured as follows.
In Section~\ref{sec:obs}, we describe the preparation of the X-ray data and the construction of a raw list of X-ray sources in the LW.
We then provide in Section~\ref{sec:methods} a brief overview of various period searching methods, with an emphasis on the advantage of the Gregory-Loredo algorithm. We also provide an outline of the algorithm's basic principles, supplementing in the Appendix the mathematical formulae in detail.
This section is closed with an evaluation of the efficiency and completeness of the Gregory-Loredo algorithm by using a large set of simulated periodic light curves. 
Section~\ref{sec:results} is devoted to the period searching results, followed by an X-ray spectral analysis of the 23 confirmed periodic sources as presented in Section~\ref{sec:spectra}.
A comparison with previous work, an attempt to classify the detected periodic sources based on their observed properties, and implications on the bulge population of CVs are addressed in Section~\ref{sec:discussion}.
A summary of this study is given in Section~\ref{sec:summary}.

\section{X-ray Data Preparation} \label{sec:obs}
\subsection{{\it Chandra} observations} \label{subsec:xdata}
The LW towards the inner Galactic bulge has been extensively observed by {\it Chandra} with its Advanced CCD Imaging Spectrometer (ACIS).
A total of 13 ACIS-I observations were taken, three in 2005 and ten in 2008, resulting in a total exposure of 982 ks.
A log of these observations is given in Table \ref{tab:obsinfo}. 
A number of previous studies have made use of all or part of these observations, which primarily focused on the identification of discrete X-ray sources and the quantification of their statistical properties \citep{2009Natur.458.1142R,2009ApJ...700.1702V,2009ApJ...706..223H,2011MNRAS.414..495R,2012MNRAS.427.1633H,2013ApJ...766...14M,2016MNRAS.462L.106W}.

\begin{table*}
\centering
\caption{{\it Chandra} observations of the Limiting Window} \label{tab:obsinfo}
\centering
\begin{tabular}{ccccccc}
\hline
\hline
ObsID & Start Time & Nominal R.A. & Nominal Decl. &  Roll angle & Exposure & Mode\\
& UT & ($\circ$) & ($\circ$) & ($\circ$) & ks & \\ 
\hline
6362 & 2005-08-19 16:15 & 267.86875 & -29.58800 & 273 & 37.7 & FAINT \\
5934 & 2005-08-22 08:16 & 267.86875 & -29.58800 & 273 & 40.5 & FAINT \\
6365 & 2005-10-25 14:55 & 267.86875 & -29.58800 & 265 & 20.7 & FAINT \\
9505 & 2008-05-07 15:29 & 267.86375 & -29.58475 & 82  & 10.7 & VFAINT \\
9855 & 2008-05-08 05:00 & 267.86375 & -29.58475 & 82  & 55.9 & VFAINT \\
9502 & 2008-07-17 15:45 & 267.86375 & -29.58475 & 281 & 164.1 & VFAINT \\
9500 & 2008-07-20 08:11 & 267.86375 & -29.58475 & 280 & 162.6 & VFAINT \\
9501 & 2008-07-23 08:13 & 267.86375 & -29.58475 & 279 & 131.0 & VFAINT \\
9854 & 2008-07-27 05:53 & 267.86375 & -29.58475 & 278 & 22.8 & VFAINT \\
9503 & 2008-07-28 17:37 & 267.86375 & -29.58475 & 275 & 102.3 & VFAINT \\
9892 & 2008-07-31 08:07 & 267.86375 & -29.58475 & 275 & 65.8 & VFAINT \\
9893 & 2008-08-01 02:44 & 267.86375 & -29.58475 & 275 & 42.2 & VFAINT \\
9504 & 2008-08-02 21:23 & 267.86375 & -29.58475 & 275 & 125.4 & VFAINT \\
\hline
\end{tabular}
\end{table*}

\begin{figure*}
\centering
\includegraphics[scale=0.8]{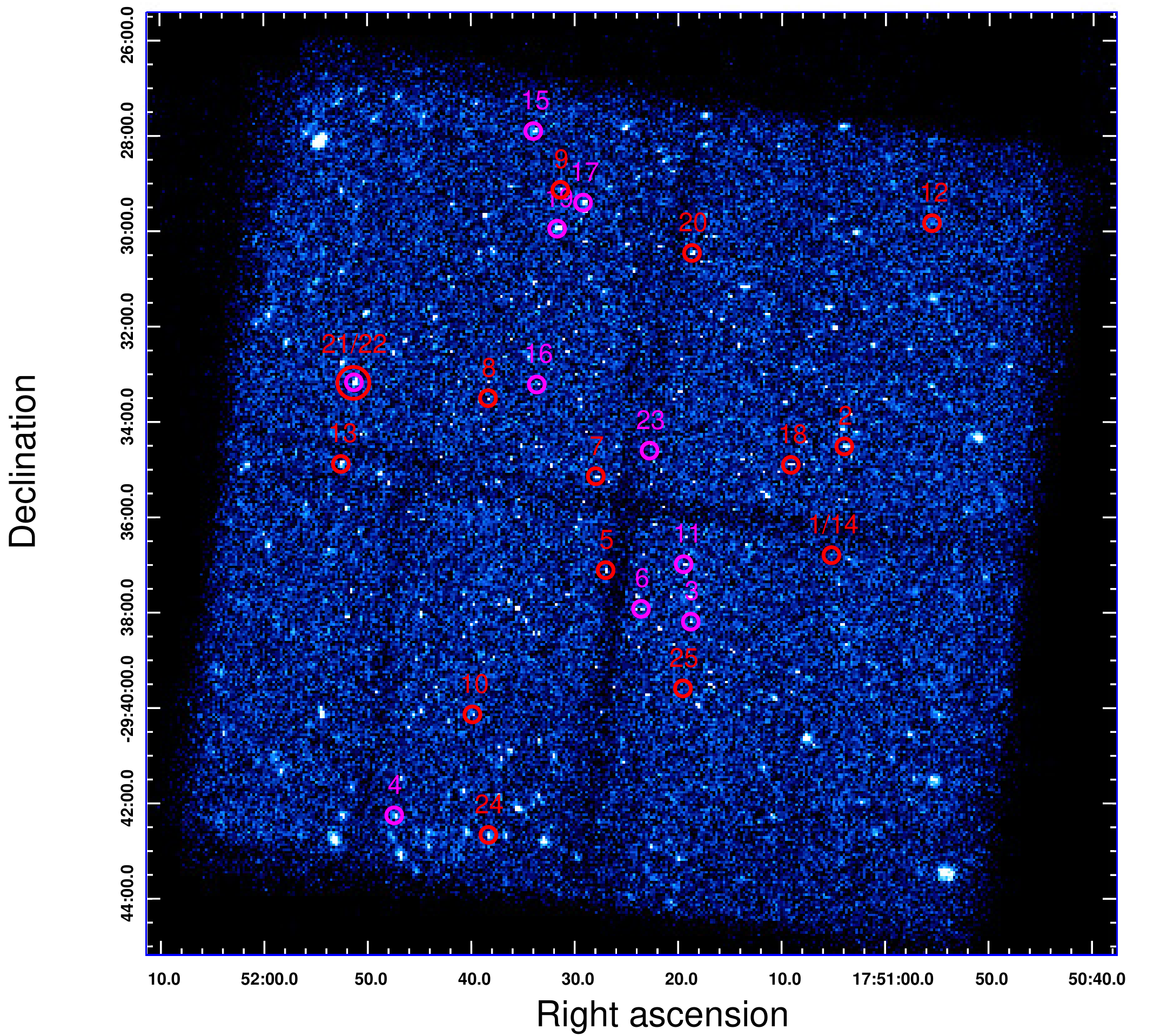}
\caption{2--8 keV counts image of the Limiting Window, combining 13 {\it Chandra}/ACIS-I observations. Locations of the 23 periodic sources are marked with colored circles ({\it magenta}: ten sources previously reported by \citealp{2012ApJ...746..165H}; {\it red}: thirteen newly discovered in this work). Source numbering is the same as in Table~\ref{tab:src}. In particular, 1/14 and 21/22 are the two sources each showing two periodic signals.}
\label{fig:FoV}
\end{figure*}

We downloaded and uniformly reprocessed the archival data with CIAO v4.10 and CALDB v4.8.1, following the standard procedure\footnote{http://cxc.harvard.edu/ciao}.
The CIAO tool \emph{reproject\_aspect} was employed to align the relative astrometry among the individual observations, by matching the centroids of commonly detected point sources. ObsID 9502, which has the longest exposure (164.1 ks), served as the reference frame.
The level 2 event file was created for each ObsID, with the arrival time of each event corrected to the Solar System barycenter (i.e., Temps Dynamique Barycentrique time) by using the CIAO tool \emph{axbary}.
We then constructed a merged event list, reprojecting all events to a common tangential point, [R.A., Decl.]=[267.86375, 29.58475].
The individual observations cover a similar field-of-view (FoV) of $\sim$70 arcmin$^2$, due to their similar aimpoints and roll angles. The FoV is illustrated in Figure~\ref{fig:FoV}, which displays the merged 2--8 keV counts image.
We have examined the light curve of each ObsID and found that the instrumental background was quiescent for the vast majority of time intervals.
Hence we preserved all the science exposures for source detection and subsequent timing analysis, taking the advantage of uninterrupted signals within each observation.  

\subsection{Source detection}\label{subsec:detect}
It is known that the LW suffers from moderate line-of-sight absorption, $N_{\rm H} \approx 7\times10^{21}{\rm~cm^{-2}}$ \citep{2011MNRAS.414..495R}, which obscures X-ray photons with energies $\lesssim$ 1 keV.
Here we focus on sources prominent in the 2--8 keV band, which are most likely CVs located in the Galactic bulge. This will also facilitate a direct comparison with the CVs found in the Nuclear Star Cluster \citep{2018ApJS..235...26Z}, the line-of-sight column density of which, $N_{\rm H} \sim 10^{23}{\rm~cm^{-2}}$, is only transparent to photons with energies $\gtrsim$2 keV. 

Source detection was performed following the procedures detailed in \citet{2018ApJS..235...26Z}.
Briefly, we first generated for each observation an exposure map as well as point-spread function (PSF) maps with enclosed count fraction (ECF) of 50\% and 90\%. 
Both the exposure and PSF maps were weighted by a fiducial spectrum, which is an absorbed bremsstrahlung with a plasma temperature of 10 keV and a column density of $N_{\rm H}=10^{22}{\rm~cm^{-2}}$, representative of the X-ray sources in the LW. 
We then reprojected the individual exposure maps to form a stacked exposure map in the same way as for the merged counts image; the PSF maps were similarly stacked, weighted by the corresponding exposure map. 
Next, we employed {\it wavdetect} to identify discrete sources in the merged 2--8 keV counts image, supplying the algorithm with the stacked exposure map and the 50\%-ECF PSF map and adopting a false-positive probability threshold of $10^{-6}$. 
This resulted in a raw list of 847 independent sources in the 2--8 keV band. 
The source centroid derived from {\it wavdetect} was refined using a maximum likelihood method that iterates over the detected counts within the 90\% enclosed counts radius (ECR).
Starting from this step we include counts detected in the 1--8 keV band to maximize the signal from potential sources in the LW. 
Then, for each ObsID, source counts were extracted from the 90\% ECR, while background counts were extracted from a concentric annulus with inner-to-outer radii of 2--4 times the 90\% ECR, excluding any pixel falling within 2 times the 90\% ECR of neighboring sources.
Crowding of X-ray sources is not a general concern for the LW, but in a few cases the source extraction region was reduced to 50\% ECR due to otherwise overlapping sources. 
The total source and background counts were obtained by summing up the individual observations. 
Photometry (i.e., net photon flux and its error) for individual sources were calculated using the CIAO tool \emph{aprates}, which takes into account the local effective exposure, background and ECF. 
We consider a {\it significant detection} for a given source in a given ObsID if the photon flux is greater than 3 times the statistical error. 
We further define for each source an inter-observation {\it variability index}, $\rm VI=S_{max}/S_{min}$, where $\rm S_{max}$ and $\rm S_{min}$ are the maximum and minimum photon fluxes among all the significant detections, respectively. This implicitly requires significant detections in at least two observations.

\section{Period Searching Method}\label{sec:methods}
In this section, we first provide our motivation of employing the Gregory-Loredo (GL) algorithm, followed by an outline of its basic principles (Section~\ref{subsec:GL}). We then describe our application of the GL algorithm to the {\it Chandra} data of the LW (Section~\ref{subsec:appli}). This is complemented by a set of simulations to evaluate the detection (in)completeness of periodic signals (Section~\ref{subsec:simulation}).  
\subsection{The Gregory-Loredo Algorithm} \label{subsec:GL}
There exists in the literature a variety of period searching methods, which can be broadly divided into three categories according to their working principles. 

The traditional method is based on Fourier transform and its power density spectra, which includes the classical Schuster periodogram \citep{1898TeMag...3...13S}, the Fourier analysis with unequally-spaced data \citep{1975Ap&SS..36..137D}, the correlation-based method \citep{1988ApJ...333..646E}, among others.

Another widely-used method seeks to fit the data with a periodic model in the frequency space, employing statistics such as least-squares residuals to define the likelihood function and then selecting the frequency that maximizes the likelihood. The famous Lomb-Scargle periodogram \citep[hereafter LS]{1976Ap&SS..39..447L,1982ApJ...263..835S} belongs to this category. Note that when adopting trigonometric functions, the least-squares method falls into the Fourier transformation category. Another variant is to replace the least-squares residuals with polynomial fits, such as that used in \citet{1996ApJ...460L.107S}.

The last category is the phase-folding method. For each trial period the time-tagged data is folded as a function of phase, and the best-fit period is found by optimizing the cost function through the frequency space. The cost function is designed to evaluate how much the phase-folded light curve deviates from constant.
Methods belonging to this category use diverse cost functions. Several widely known examples are the Epoch Folding (EF) algorithm \citep{1983ApJ...266..160L}, the Phase Dispersion Minimization \citep{1978ApJ...224..953S}, and the GL algorithm \citep{1992ApJ...398..146G}.

In X-ray studies, the detection of periodic signals is often involved with irregularly and sparsely sampled data. 
When working in frequency space, such a sampling can lead to spurious signals and heavy contamination to the real signal. Phase-folding methods, on the other hand, can minimize the effect of irregular data since the cost function excludes the dead time and can compensate for observation gaps.
Moreover, the number of detected source counts is often only moderate. While one can in principle apply binning to create photometric light curves, it comes at the price of potentially losing temporal information. Phase-folding methods, on the other hand, directly handle individual events and thus can maximally incorporate the temporal information.

Most X-ray sources in the LW share the characteristics of irregular sampling and limited source counts. 
Therefore, it is appropriate to employ the GL algorithm, which applies the Bayesian probability theorem to the phase-folded light curve, to search for periodic signals in the LW sources. We provide a brief summary in mathematical form of Bayes's theorem and the GL algorithm in Appendix \ref{GL}. 
The key of this algorithm is the multiplicity of the phase distribution of events (i.e., detected counts),
\begin{equation}\label{multi}
W_m(\omega, \phi)={{N!}\over{n_1!\; n_2!\; n_3!\cdots {n_m}!}}.
\end{equation}
Here $N$ represents the total number of counts of a given source, 
$n_i(\omega, \phi)$ is the number of counts falling into the $i$th of $m$ phase bins, given the frequency $\omega$ and phase $\phi$, satisfying $\sum\limits_{i=1}^{m}n_i(\omega, \phi)=N$. 
The multiplicity is the number of ways that the binned distribution could have arisen by chance. It can be easily shown that the more the values of $n_i$ differ from each other, the smaller the multiplicity. In other words, the more the stepwise model defined by the $m$ phase bins deviates from constant, the more likely there exists a periodic signal, the probability of which is inversely proportional to the multiplicity.  

In general, the GL algorithm takes the following steps:

(i) Compute the multiplicity for all sets of $(m,\omega, \phi)$ (Eqn.~\ref{multi}). In this work, the highest value of $m$ is set to be 12.

(ii) Given $m$, integrate over the $(\omega, \phi)$ space and calculate the so-called ``odds ratio'' using Bayes's theorem (Eqn.~\ref{A16}). The ``odds ratio'' determines the ratio of probabilities between a periodic model and a non-periodic (constant) model. The range of $\omega$ depends on the source of interest and is further addressed in Section~\ref{subsec:appli}.

(iii) Sum up the normalized odds ratios of each $m$ to determine the probability of a periodic signal (Eqn.~\ref{A20}).
If this probability exceeds a predefined threshold (default at 90\%), a periodic signal is favored. 

(iv) Finally, compare all the odds ratios integrated over the $\phi$ space (0--2$\pi$), finding the value of $\omega$ with the highest odds ratio, which then gives the period $P=2{\pi}/\omega$ (Eqn.~\ref{A21}).

In reality, due to random flux  variability that typically exist in accretion-powered sources, the assumption of a constant baseline flux, as designed by the GL algorithm, may not be strictly satisfied. However, as the variability index (Section~\ref{subsec:detect}) indicates, the majority of LW sources exhibit only moderate long-term flux variation, and because of their intrinsic faintness, their short-term flux variation is dominated by statistical fluctuations, i.e., the Poisson process that is integrated in the GL algorithm.

\subsection{Application to the LW}\label{subsec:appli}
We apply the GL algorithm to search for periodic signals in the LW sources. 
For a given source, the 1--8 keV counts within the 90\% ECR of individual ACIS-I observations are extracted to form a time series.  
In addition, we supply for each source the information of ``epoch'', i.e., the start time and end time of each ObsID. This information is used to compensate for the uneven distribution of exposure over the phase bins (see Eqn.~\ref{A17} for the exact treatment). 
Since the GL algorithm determines the probability of a periodic signal against a constant light curve, there is no need to separately account for the background level, which is absorbed into the constant. 
Nevertheless, we have measured the local background (Section~\ref{subsec:detect}) for each periodic source as a consistency check (see Section~\ref{sec:results}). 

As mentioned in Section~\ref{subsec:GL}, the GL algorithm folds the time series at a trial frequency (or period). 
In practice, the resolution and range of frequency must be compromised between efficiency and computational power. 
Thus we restrict our analysis on three period ranges: (300, 3000), (3000, 10000) and (10000, 50000) sec, with a frequency resolution of $10^{-7}$, $10^{-8}$ and $10^{-9}$ Hz, respectively. 
After finding a tentative period in a certain period range, the GL algorithm is run a second time, but with a narrow interval around the identified period excluded, to ensure that a possible second period within the same range will not be missed.
The period ranges are chosen based on the expectation that most, if not all, detectable periodic X-ray sources in the LW should be CVs. 
The orbital period distribution of CVs is known to exhibit a minimum at $\sim$82 minutes, a gap between 2--3 hours, and a maximum around 12 hours 
\citep{2011ApJS..194...28K}.
The second and third period ranges well cover these characteristic periods, whereas the first range probes the spin period of fast rotating IPs.
Given the timespan of $\sim10^8$ sec between the first and last ACIS-I observations, the chosen frequency resolutions are optimal for an efficient search of periodic signals. 

\subsection{Detection completeness}\label{subsec:simulation}
For a given period searching algorithm, the detection rate depends  mainly on the number of observed counts, the intrinsic shape of the light curve, as well as the observing cadence. 
To quantify the detection rate and hence gain insight on the nature of the periodic sources in the LW, we perform simulations following the merit of \citet{1998ApJ...498..666C}. 
Two functional forms of light curve are considered: a sinusoidal function and a piecewise function. While these are admittedly idealized shapes, they can represent realistic light curves, e.g., the former resulted from rotational modulation and the latter due to eclipse. 
A sinusoidal light curve follows,
\begin{equation}
\lambda (t)=\lambda_0[1+A_{0}{\rm sin}(\omega t+\phi)], 
\label{eqn:sin}
\end{equation}
where $\omega = 2{\pi}/P$, $A_0$ is the relative amplitude of variation, and $\lambda_0$ is the mean count rate which may include contribution from a constant background. The phase $\phi$ can be arbitrarily set at zero.
For a direct comparison with observations, we relate $\lambda_0$ to the total number of counts, $C = \lambda_0 T_{\rm exp}$, where $T_{\rm exp}$ is the exposure time. This holds since $T_{\rm exp}$ is much longer than the modulation period ($P$). 
The simulations are run with a selected number of parameters due to constraints in computational power. 
Specifically, we adopt $C$=50, 100, 200, 300, 400 and 500, $A_0$=0.5, 0.6, 0.7, 0.8 and 0.9, and $P$=554, 5540 and 45540 sec, resulting in a total of 6x5x3=90 combinations. 
The chosen periods are representative of the actually searched ranges (Section~\ref{subsec:appli}), whereas
the adopted total counts well sample the range of observed counts in the LW sources.

We thus simulate 100 light curves for each combination of parameters.
 The photon arrival time, taking into account the exact start and end times of the 13 ACIS-I observations, is simulated with the Poisson process according to the time-dependent count rate of Eqn.~\ref{eqn:sin}, and is further randomly modified by an amount of 3.2 sec to mimic the effect of ACIS frame time.
The simulated light curve is then searched for a periodic signal using the GL algorithm in the same manner as for the real data (Section~\ref{subsec:appli}).
We count a valid detection if the identified period has a detection probability greater than 90\% and its value is consistent with the input period to within 1\%. 
We notice that in a small fraction of simulated light curves of the shortest input period (554 sec), the second harmonics (i.e., 2 times the true period) is detected with an even higher probability than the true period. We also consider such cases a valid detection as long as the true period itself fulfills the above criteria.

\begin{figure}
\begin{minipage}[b]{0.45\textwidth}
\includegraphics[width=\textwidth]{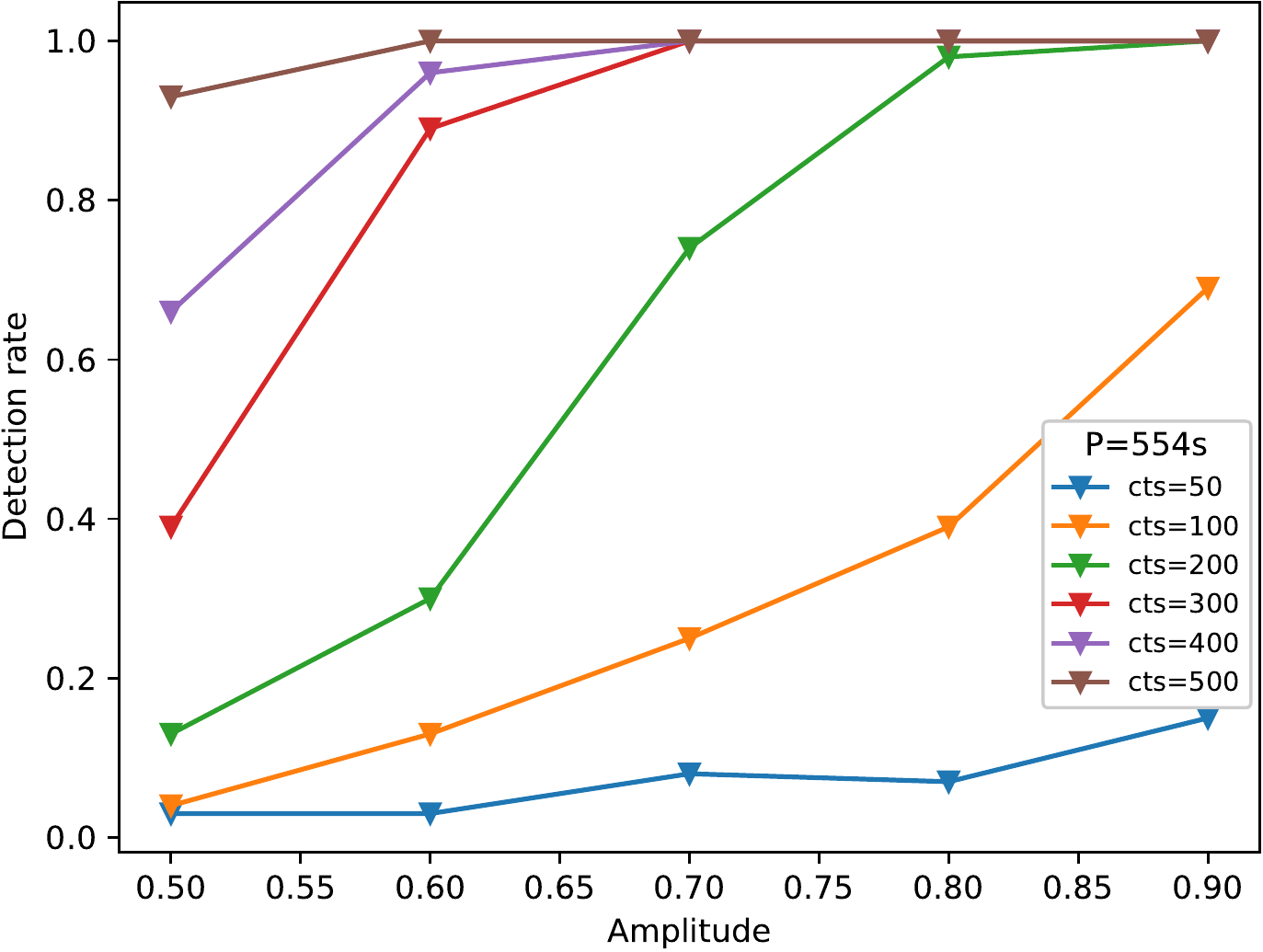}
\includegraphics[width=\textwidth]{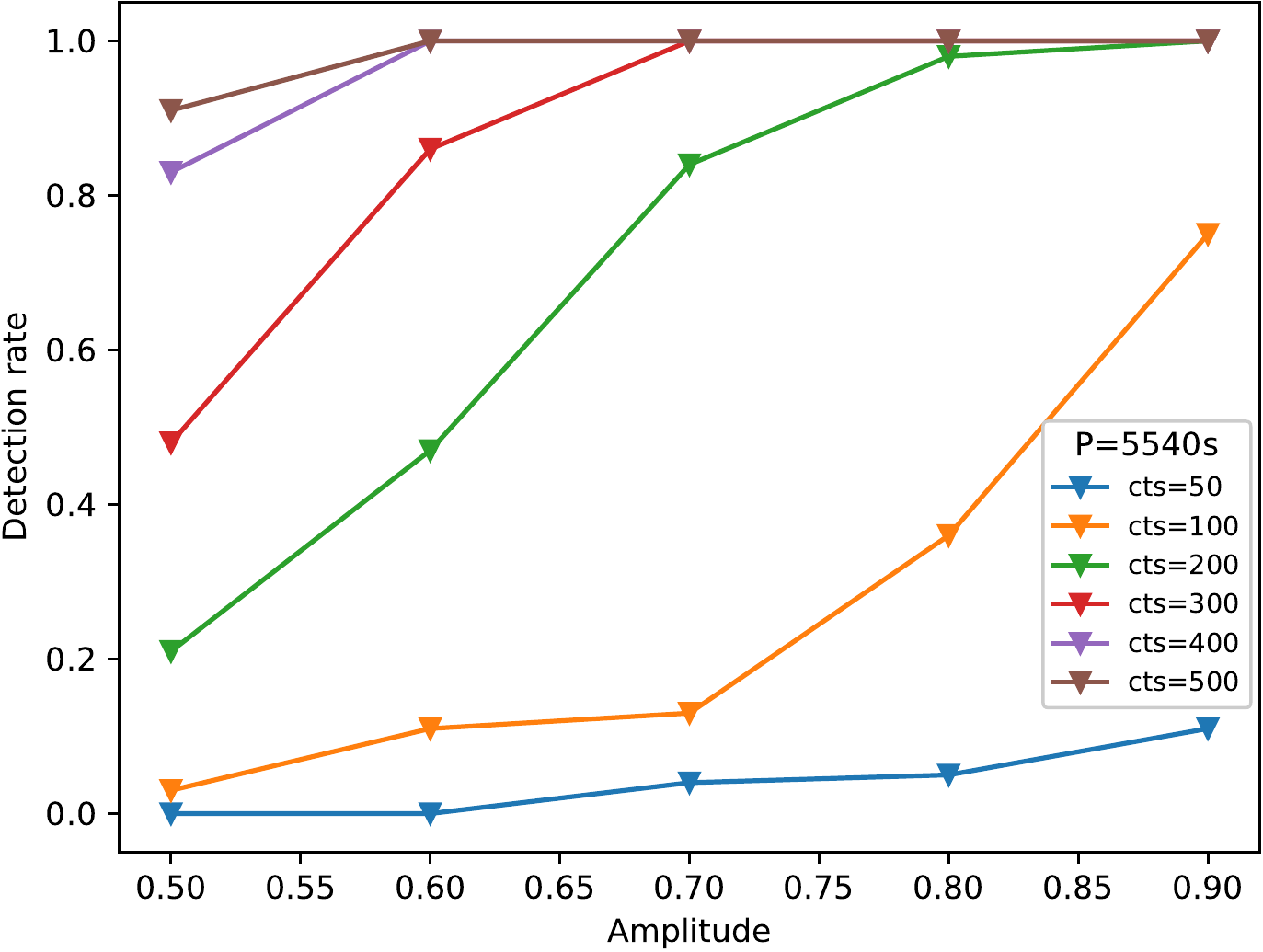}
\includegraphics[width=\textwidth]{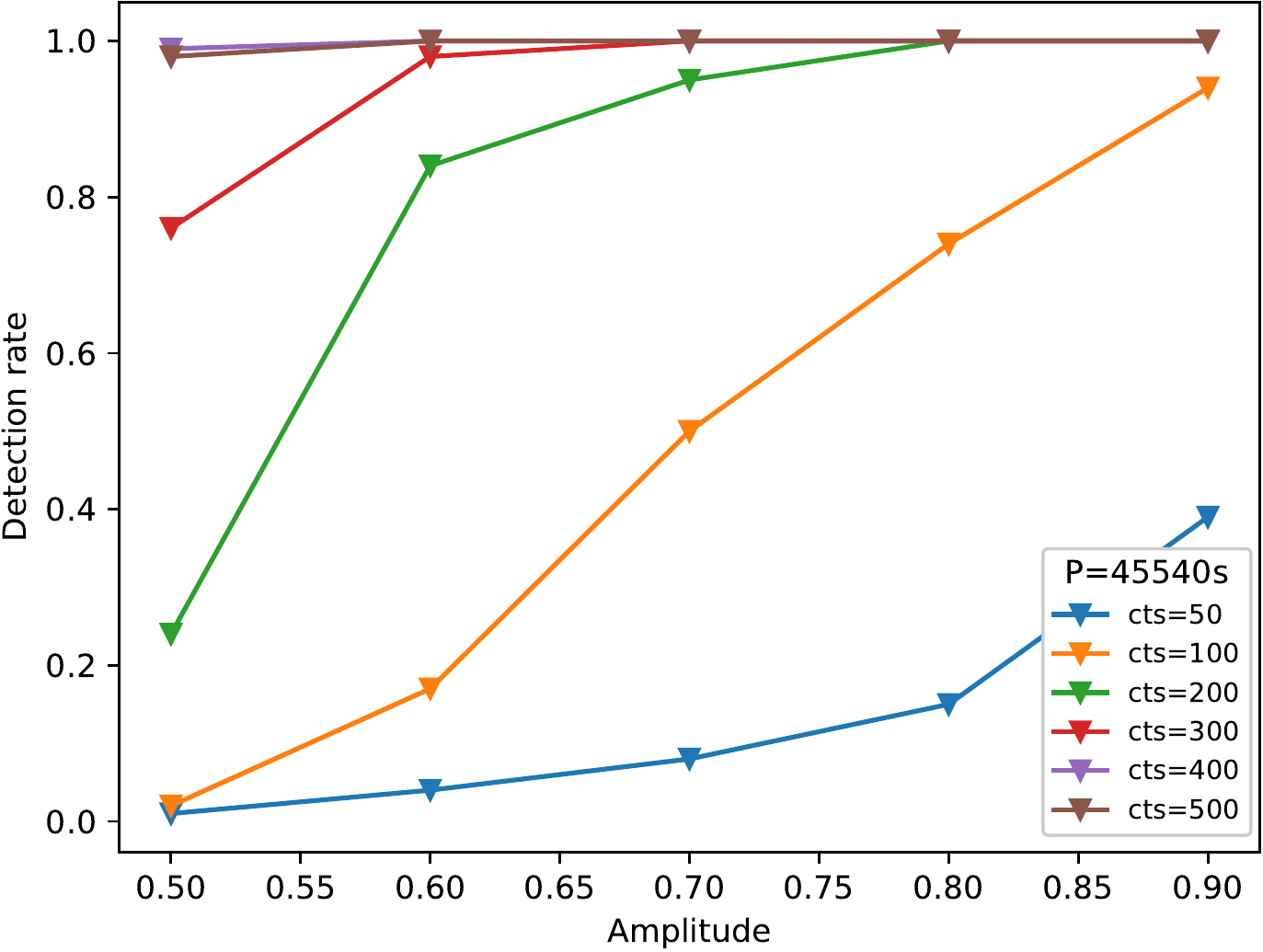}
\end{minipage}
\caption{Detection rates as a function of relative variation amplitude, based on simulated sinusoidal light curves. The top, middle and bottom panels are for modulation period of 554, 5540 and 45540 sec, respectively. The different colored symbols and lines represent different values of total counts, as labeled.
\label{fig:detection}}
\end{figure}

The detection rate for a given combination of parameters is taken to be the fraction of the 100 simulated light curves having a valid detection. The top, middle and bottom panels of Figure \ref{fig:detection} show the result of the three test periods, respectively, in which several trends are apparent. 
First and intuitively, for a given period and amplitude, higher total counts would lead to higher detection rates, while for given total counts, a higher amplitude also leads to a higher detection rate. The simulation results confirm these expectations.  
Second, for the same total counts and amplitude, the detection rate is generally higher for a longer period.  
This can be understood as due to a statistical behavior in the 
multiplicity (Eqn.~\ref{multi}), which has a lower value for a longer period. This holds for both the sinusoidal and piecewise light curves. 
Lastly, for total counts of 50, the detection rate is almost always below 10\% regardless of the period and amplitude. 
 
\begin{figure}
\centering
\includegraphics[scale=0.61]{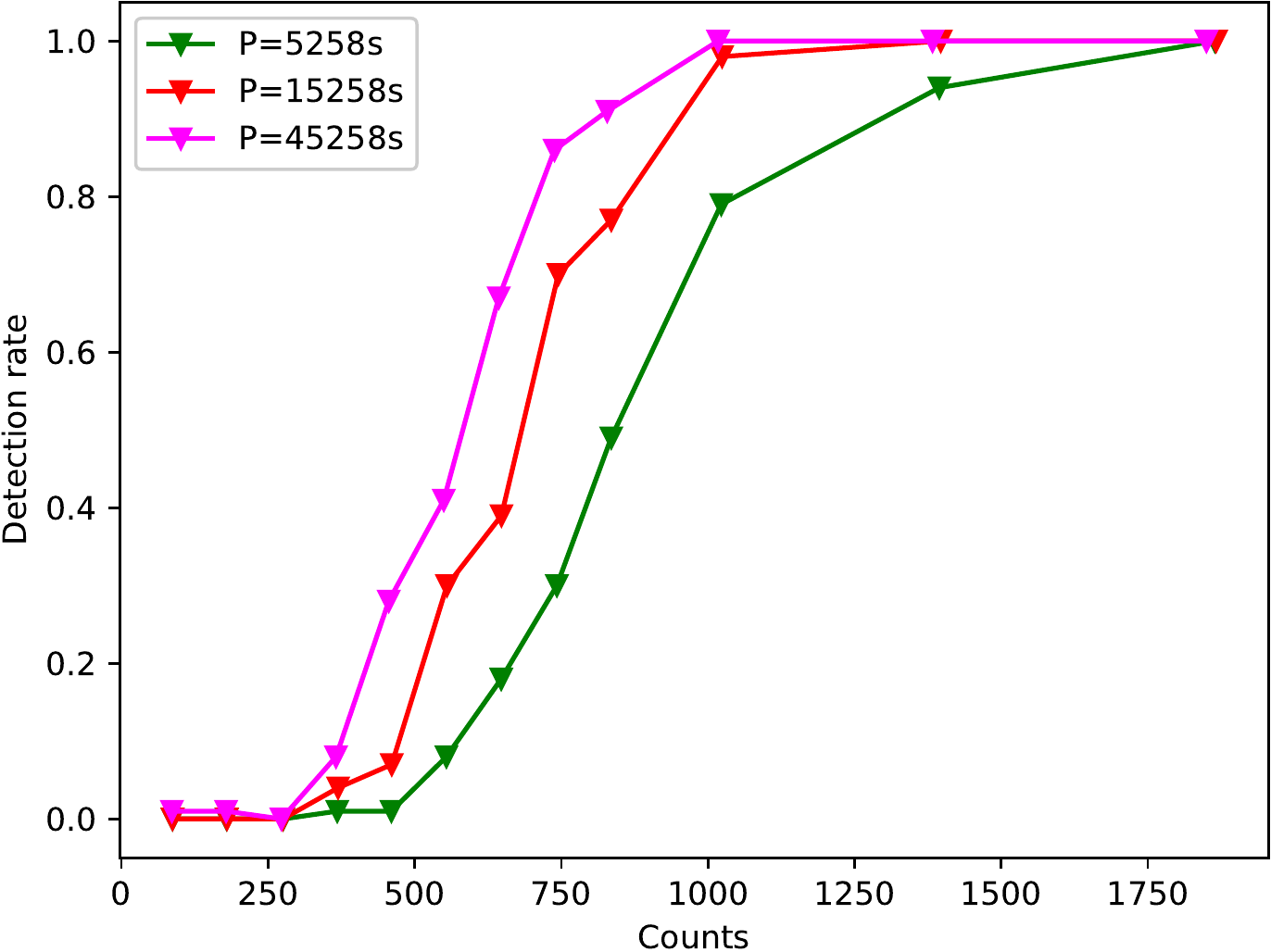}
\caption{Detection rates as a function of total counts, based on simulated piecewise light curves. Different colored symbols and lines represent different periods, as labeled.}\label{fig:eclipse}
\end{figure}

The piecewise function, which mimics an eclipse against an otherwise constant flux, takes the form of
\begin{equation}
\lambda(t)=
\begin{cases}
\lambda_0 & \text{$\phi(t) \in[0,(1-w)\pi)\cup ((1+w)\pi,2\pi]$},\\
f\lambda_0 & \text{$\phi(t) \in[(1-w)\pi,(1+w)\pi]$},
\end{cases}	
\end{equation}
where $w$ accounts for the eclipse width (duration) in phase space, and $f$ characterizes the relative depth of the eclipse ($0\leq f \leq 1$; $f = 0$ corresponds to total eclipse). Here the mid-eclipse is assumed to occur at $\phi = \pi$. 
Again, $\lambda_0$ can be related to the total counts as $C=[1-(1-f)w]\lambda_0T_{\rm exp}$.
We set $f=0.1$ and $w=0.1$ in our simulations, which are not atypical of eclipsing CVs. 
We test three values of the period, $P$=5258, 15258 and 45258 sec and adopt trial count rates $\lambda_0 = 1, 2, 3, 4, 5, 6, 7, 8, 9, 10, 15$ and $20\times10^{-4}{\rm~cts~s^{-1}}$.
For each combination of parameters, 100 simulated light curves are again generated and fed to the GL algorithm. 
The resultant detection rate is shown in Figure~\ref{fig:eclipse}. 
As expected, the detection rate is generally higher for a longer period.
It can also be seen that for total counts below $\sim$300, the detection rate is $\lesssim$10\% regardless of the period; only when total counts exceed $\sim$1500, the detection rate becomes 100\% for all test periods. 
Since only a few sources in the raw list have total counts more than 1500, we expect a relatively low detection rate of eclipsing sources in the LW.  
 
We have also run simulations to estimate the rate of false detection, which refers to the detection of a periodic signal from an intrinsically non-periodic (e.g., constant) light curve. 
Considering that most LW sources can have a low-amplitude long-term variation, we approximate a non-periodic light curve with a sinusoidal function with $P$=5 yr and $A_0$=0.5. 
Simulated light curves are generated with totals counts from 50 to 5000 and fed to the GL algorithm. No false detection of periodic signals, at any value between 300--50000 sec, is found. 
Therefore, it seems safe to conclude that the false detection rate is negligible for the GL algorithm applied to the LW data. 

\section{Period Searching Results}\label{sec:results}
According to the simulations in Section~\ref{subsec:simulation}, a periodic signal is hard to detect in LW sources with total counts $C<100$, even in the case of high variation amplitudes.  
Therefore, we restrict our period searching to sources with 1--8 keV total counts $C \geq 100$, which include 667 of the 847 sources in the raw list. Among these 667 sources, 25\% (4\%) have $C \geq$ 500 (1000).

Adopting a probability threshold of 90\%, we initially find 48 tentative periodic signals from the three searched period ranges. However, it is necessary to filter spurious detections, which may be caused by several effects:

(i) By design, the ACIS is dithered to distribute photons over more CCD pixels to avoid pile-up and to fill CCD gaps. The dither period is 706.96 s in pitch and 999.96 s in yaw {\footnote{https://cxc.harvard.edu/ciao4.4/why/dither.html}}.
Any signal detected at these two periods and their harmonics are thus excluded. These are mostly found in sources located close to CCD gaps, where dithering significantly reduces the number of detected counts in a periodic fashion. 

A related concern is how dithering would affect the detection of genuine periods.  
We run simulations to test this effect. Specifically, we generate simulated sinusoidal light curves with $P$ = 5072.97 sec and $C$ = 293. 
This choice is motivated by one particular periodic source (\#5 in Table~\ref{tab:src}), which has a fractional detector coverage of 0.74 (in other words, 26\% of the intrinsic flux is lost due to dithering into the CCD gap), the lowest among all valid periodic sources. The dithering effect is mimicked by artificially removing the simulated counts according to a probability distribution calculated by the CIAO tool \emph{dither\_region}. 
No difference is found in the resultant detection rate, compared to that without dithering. 

(ii) In certain sources, multiple detections can be caused by second and third harmonics of the same intrinsic signal (i.e., 2 and 3 times the true period). These can be easily identified by the sign of double-peak or triple-peak in the phase-folded light curve, provided that the intrinsic signal has a single-peaked structure. 

However, the intrinsic structure might be double-peaked, in which case it is more difficult to distinguish between the true period and the sub-harmonic (i.e., half the true period, same below), in particular when the two peaks have a similar strength and width.
In reality, the double-peaked shape can occur in IPs producing hard X-rays near both magnetic poles (see Section~\ref{subsec:class} for more discussions). When viewed from certain angle, the two poles alternatively drift across the front side of the WD, producing the double-peaked X-ray light curve. The modulation period in this case must be the spin period of the WD, typically under one hour. The few tentative detections showing a double-peaked light curve all have a corresponding period longer than 3 hours, far beyond the empirical range of spin periods of IPs. 
These are probably second harmonics rather than a genuine spin period. 
Hence we assume that there is no sub-harmonic in the LW sources and always take the lowest period as the true period. This assumption is supported by our extensive simulations presented in Section~\ref{subsec:simulation}, in which no sub-harmonic (the true period is known before hand) is found.  

(iii) Strong flux variations or outbursts occupying one particular observation can also cause a fake periodic signature. This is because the GL algorithm, which analyzes the phase-folded light curve, can be fooled if there were too many photons found in a single observation, producing excess in certain phase bins. In this case the algorithm may ``think'' there exists a period especially in the range of (10000, 50000) sec. Among the sources with tentative periods, four exhibit a variability index VI$>10$, indicating strong variations. We thus reanalyze their light curves using two subsets of observations: those covering only the outburst and those excluding the outburst. For three of them, the tentative period cannot be recovered in either subset and thus is probably a fake signal. The remaining source is retained since its period can be recovered in both the outbursting and quiescent subsets.

The above filtering thus results in 25 valid periodic signals in 23 sources.
Among them, 10 signals were previously reported by \citet{2012ApJ...746..165H} and are confirmed here with the GL algorithm, while the remaining 15 periods are new discoveries (a comparison between our work and \citet{2012ApJ...746..165H} is further addressed in Section~\ref{subsec:compare}).

The basic information of these periodic sources are listed in Table~\ref{tab:src}, sorted by the order of increasing period. The source locations are marked in Figure~\ref{fig:FoV}.

\begin{figure*}
\begin{minipage}[t]{0.45\textwidth}
\includegraphics[width=\textwidth]{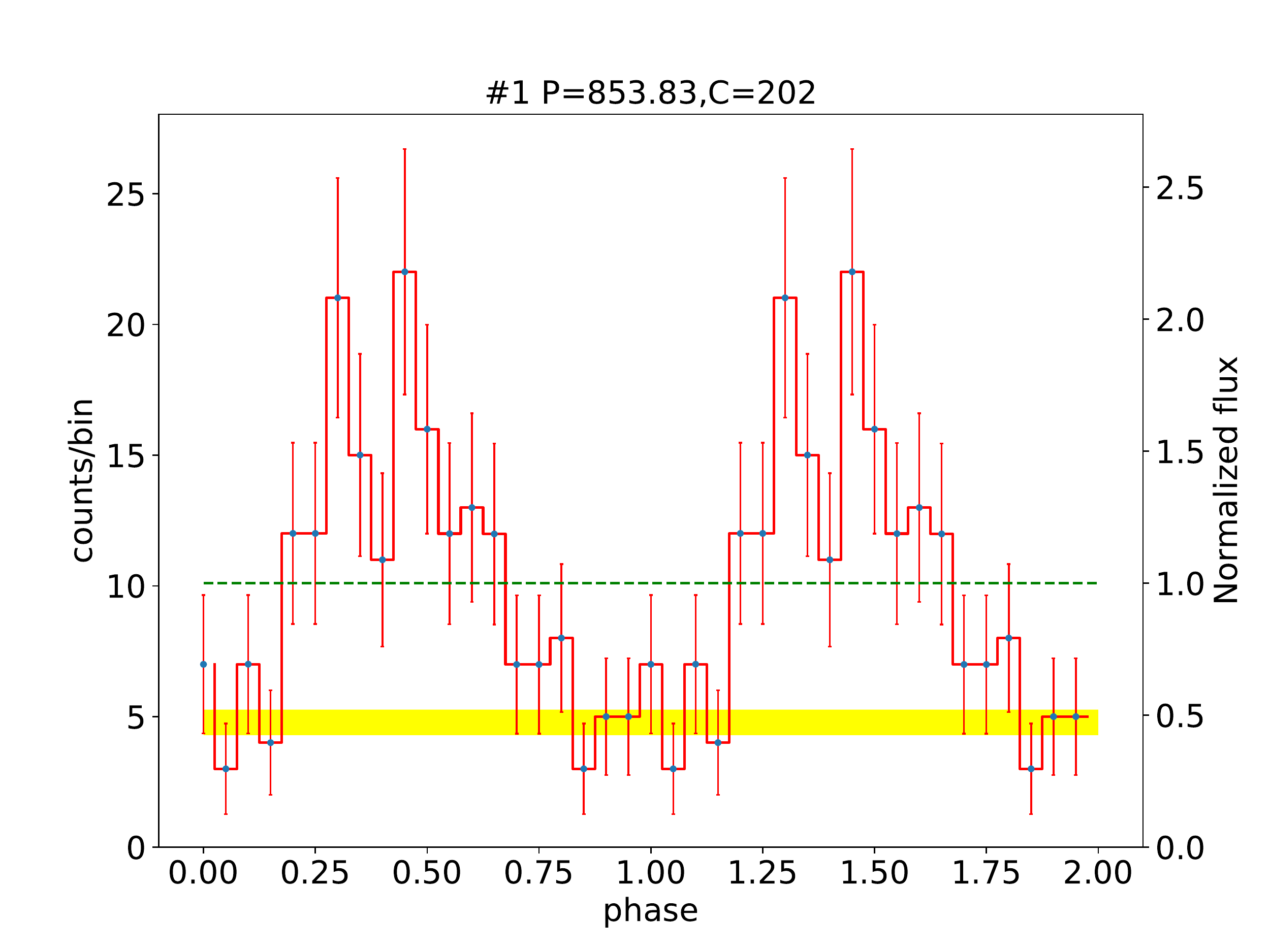}
\includegraphics[width=\textwidth]{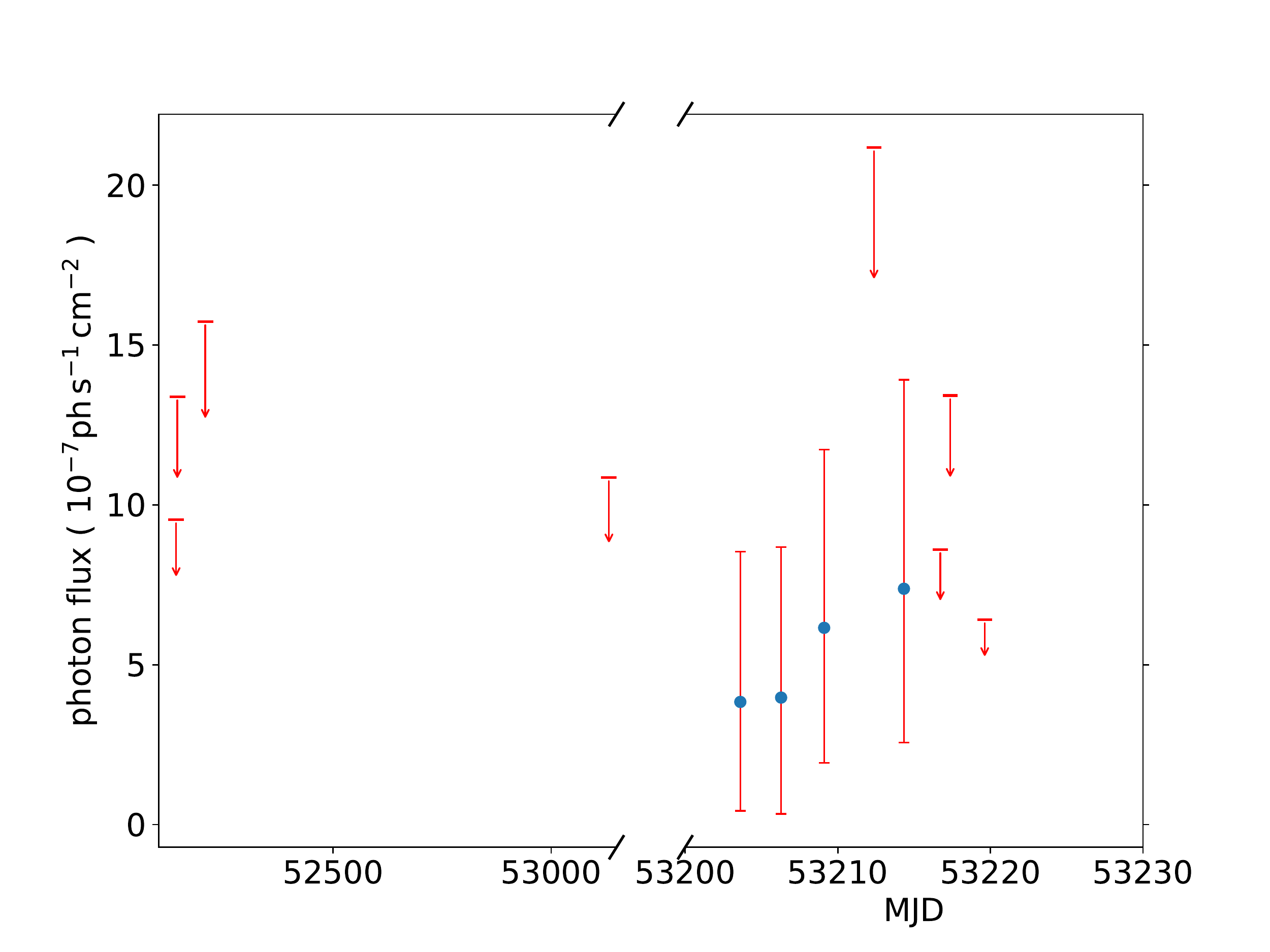}
\end{minipage}
\begin{minipage}[t]{0.45\textwidth}
\includegraphics[width=\textwidth]{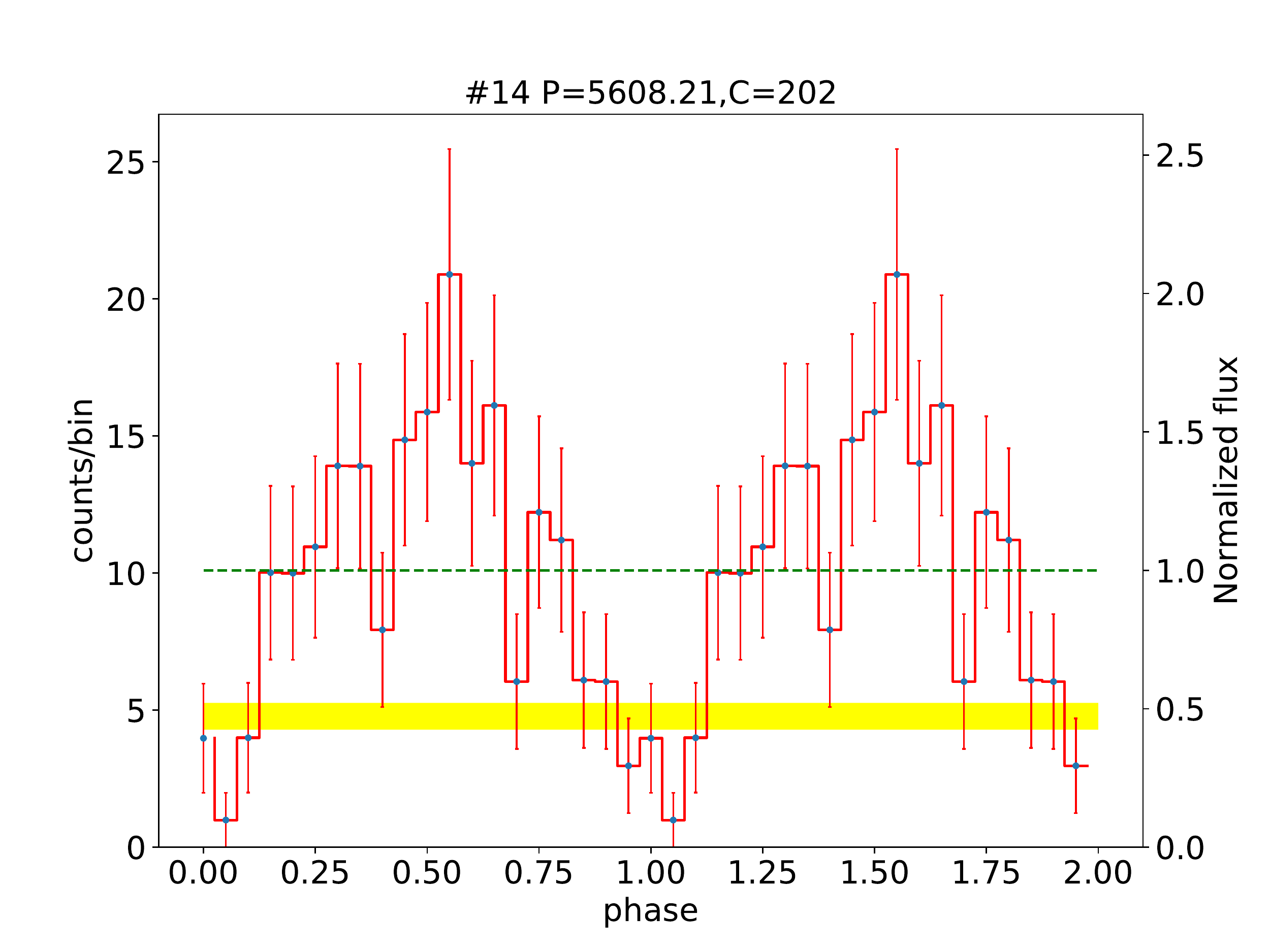}
\includegraphics[width=1.02\textwidth]{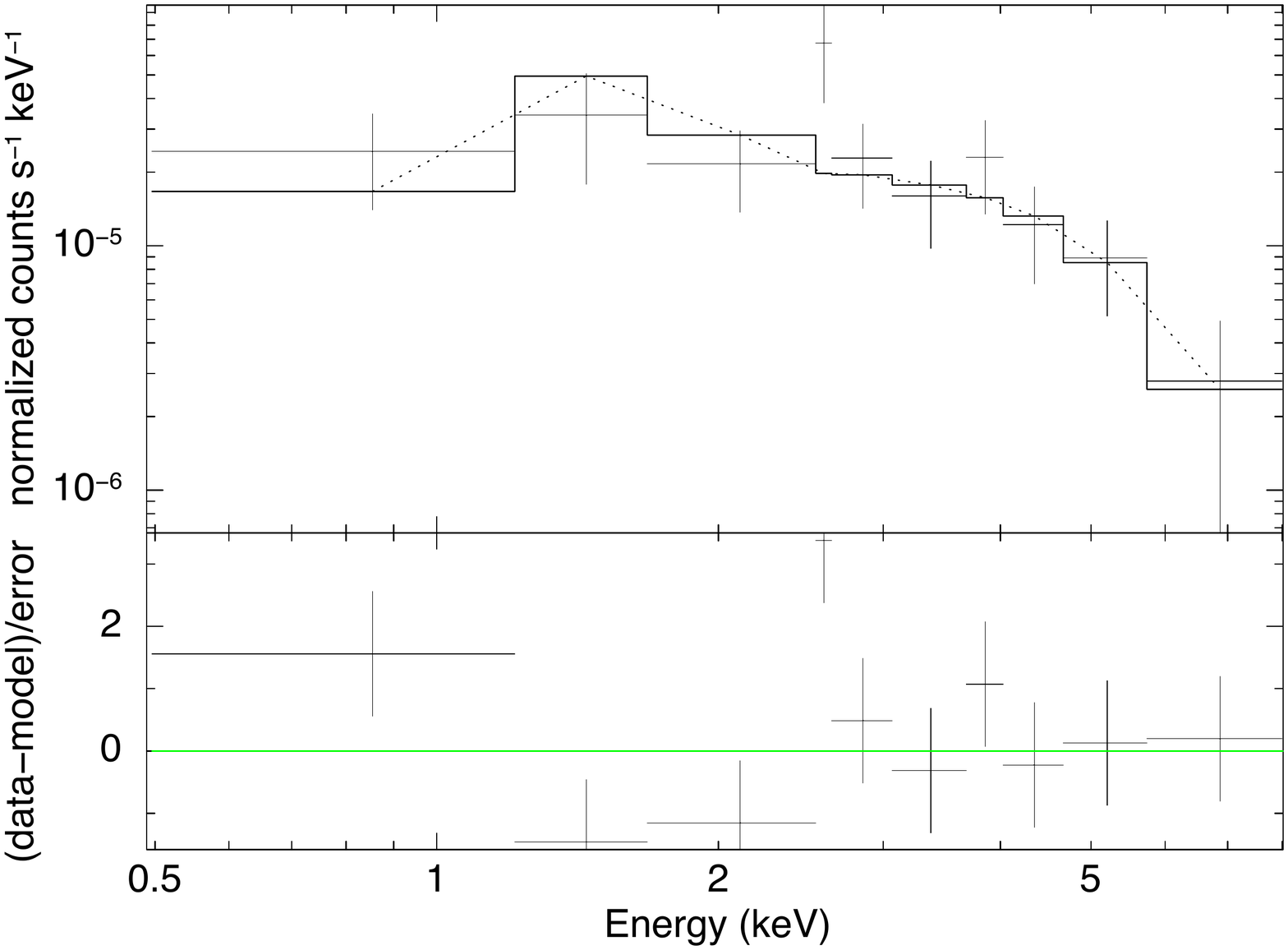}
\end{minipage}
\caption{{\it Upper panels}: The 1--8 keV phase-folded light curves of source \#1/\#14 at the two modulation periods.
The green dashed line represents the mean count rate, whereas the yellow strip represents the local background, the width of which represents 1\,$\sigma$ Poisson error.
{\it Lower left}: the 1--8 keV long-term, inter-observation light curve. Arrows represent 3\,$\sigma$ upper limits. 
{\it Lower right}: Source spectrum and the best-fit model. See text for details.}
\label{fig:pCV_sample_1}
\end{figure*}

\begin{figure*}
\begin{minipage}[t]{0.45\textwidth}
\includegraphics[width=\textwidth]{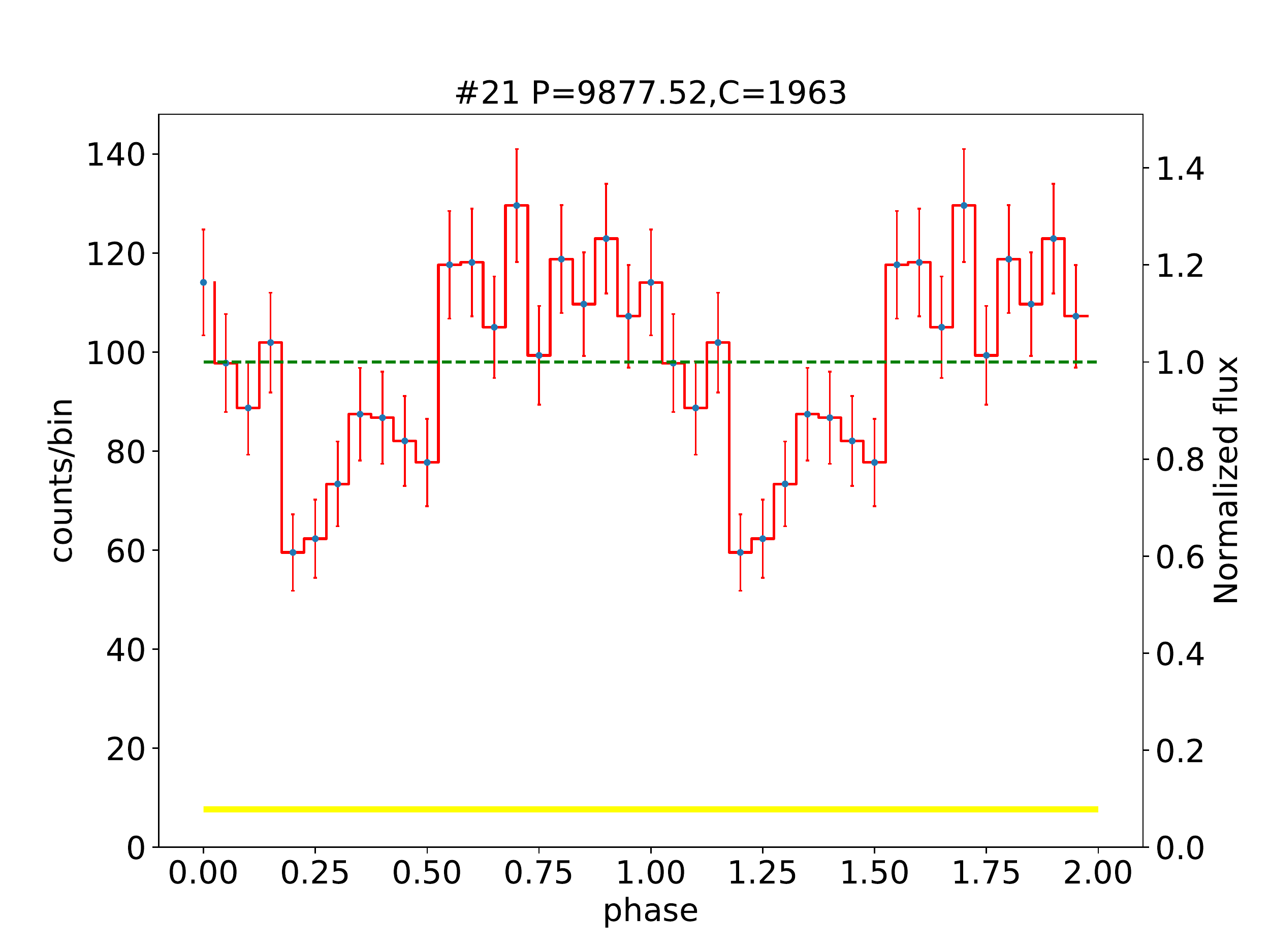}
\includegraphics[width=\textwidth]{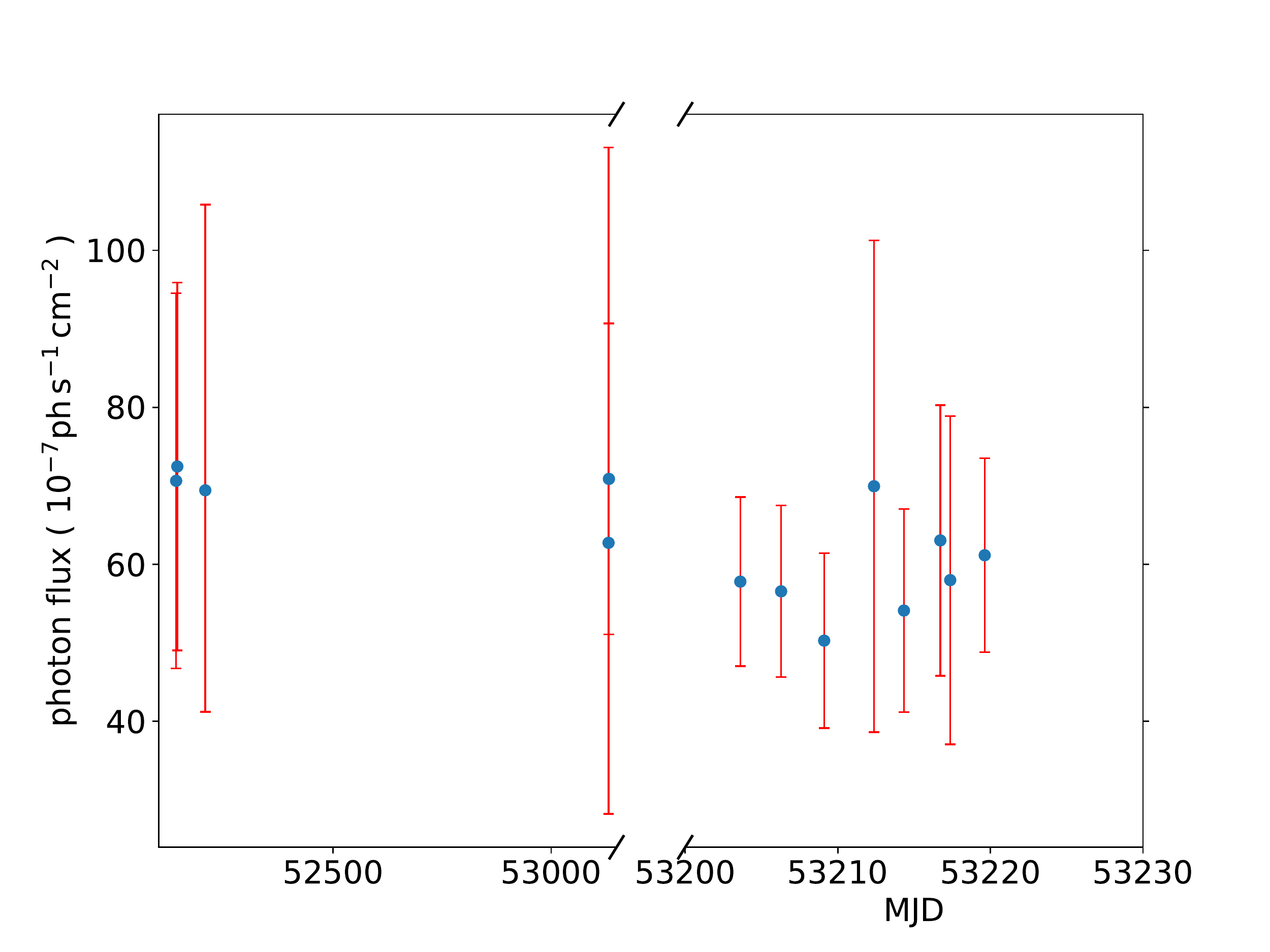}
\end{minipage}
\begin{minipage}[t]{0.45\textwidth}
\includegraphics[width=\textwidth]{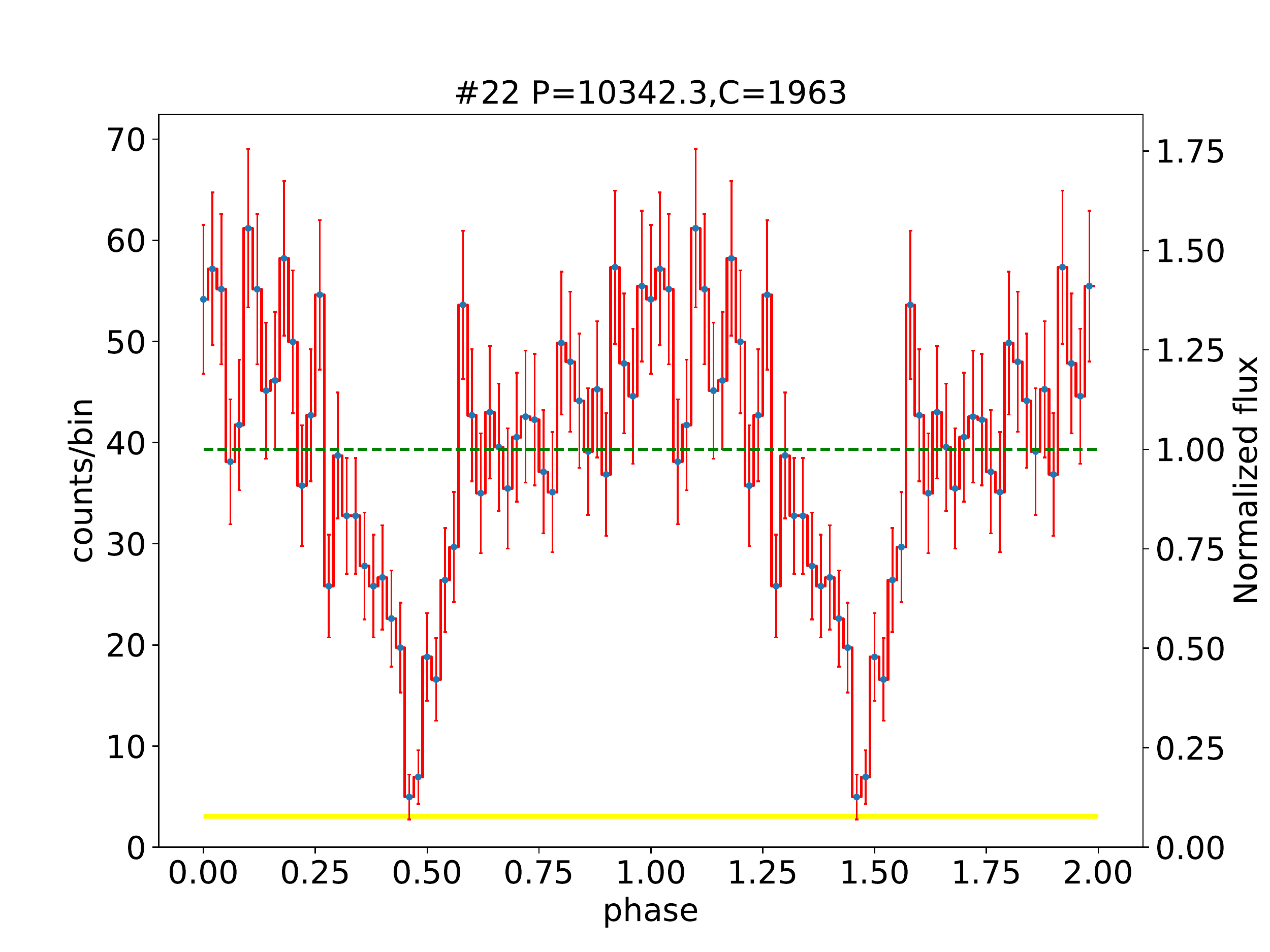}
\includegraphics[width=1.02\textwidth]{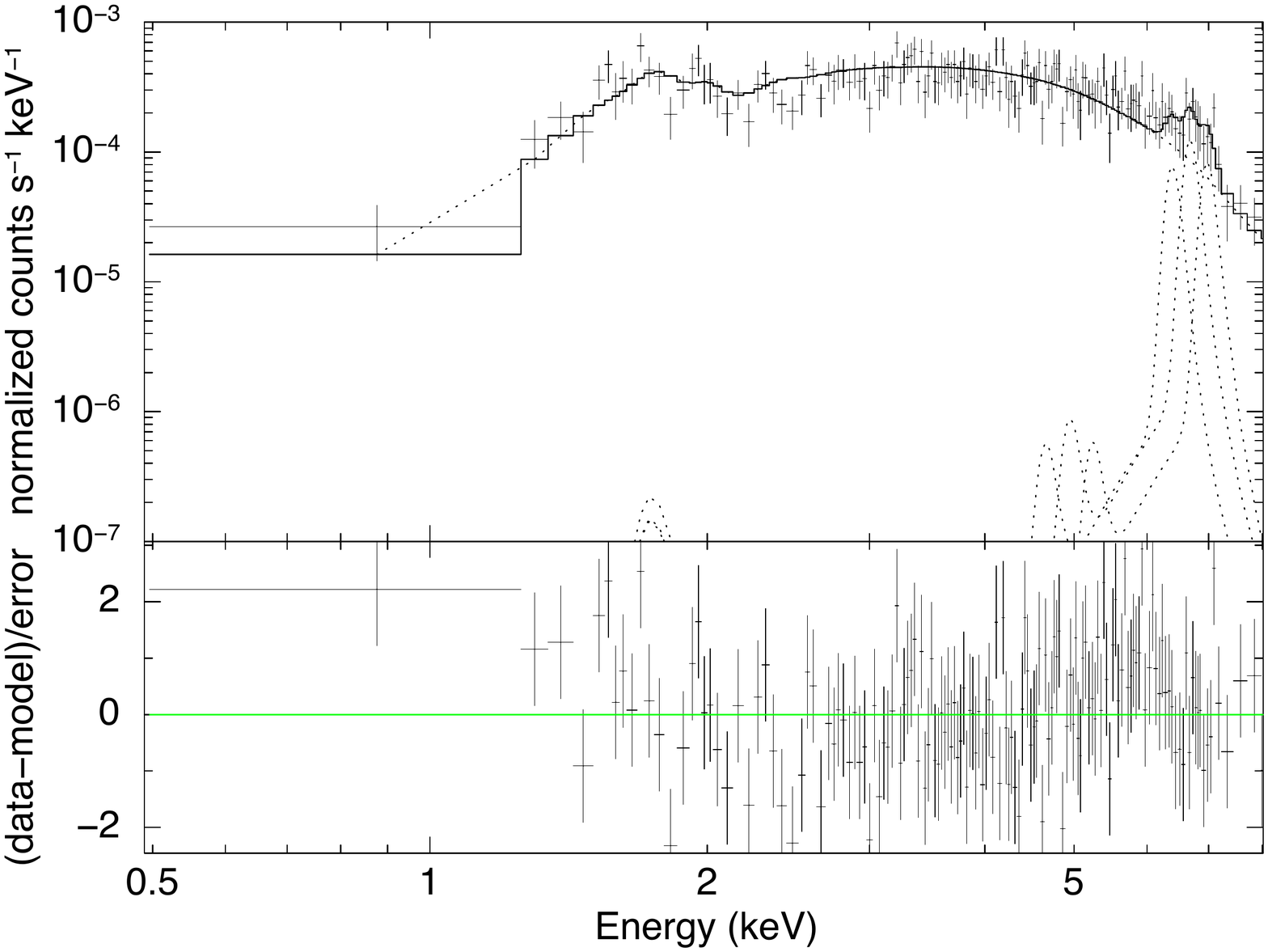}
\end{minipage}
\caption{Similar to Figure~\ref{fig:pCV_sample_1}, but for source \#21/\#22.}
\label{fig:pCV_sample_2}
\end{figure*}

There are two sources each exhibiting dual periods, hence we have assigned each of them two IDs: \#1/\#14 and \#21/\#22.  
The phase-folded light curves at the two modulation periods are shown for source \#1/\#14 in the upper panels of  Figure~\ref{fig:pCV_sample_1} and for source \#21/\#22 in the upper panels of Figure~\ref{fig:pCV_sample_2}. 
The number of phase bins, between 20 to 50, is chosen to optimally display substructures in the light curve. While the GL algorithm does not rely on quantifying the local background, for comparison we plot in these panels the estimated background level (yellow strip, the width of which represents 1\,$\sigma$ Poisson error). 
We defer discussions on the light curve shape, which contains important information on the nature of the periodic source, to Section~\ref{subsec:class}. 
 
The phase-folded light curves are complemented by the long-term, inter-observation light curve, shown in the lower left panel of Figures~\ref{fig:pCV_sample_1} and \ref{fig:pCV_sample_2}, 
and by the source spectrum (see Section~\ref{sec:spectra}),
shown in the lower right panel of Figures~\ref{fig:pCV_sample_1} and \ref{fig:pCV_sample_2}. 
Similar figures of the remaining 21 sources are presented in Appendix~\ref{appen:fig}.

\begin{table*}
\centering
\begin{threeparttable}
\caption{Basic information of the periodic X-ray sources in the Limiting Window \label{tab:src}}
\begin{tabular}{lcccccccccccc}
\hline
\hline
ID& R.A. & Decl. & Period & Prob. & $C$ & $C_{\rm B}$ & VI & H-ID  & Harmonics & Class 
\\
LW & $\circ$ & $\circ$ & s & &  &  & \% & &
\\ 
(1) & (2) & (3) & (4) & (5) & (6) & (7) & (8) & (9) & (10)  & (11)
\\
\hline
1$^\dag$ & 267.77173 &	-29.61332 & 853.83 & 0.90277 & 202 & 95.5 & 1.92  &- &- 
& IP
\\
2 & 267.76657 &	-29.57529 & 3820.83 & 0.99222 & 902 &116.4 & 1.94 &-  &- & IP?
\\
3 & 267.82829 &	-29.63660 & 4728.90 & 1.00000  & 394 & 58.3 & 2.13  & H6 & \text{Third} & IP?
\\
4 & 267.94766 &	-29.70427 & 4886.79 & 0.99994  & 784  & 413.2& 1.90 & H8  &- 
& polar?
\\
5 & 267.86255 &	-29.61859 & 5072.97 & 0.99933 & 293 & 12.9 &  2.34 & - &\text{Second} & polar?
\\
6 & 267.84831 &	-29.63212 & 5130.57 & 1.00000  & 437 & 47.2 & 2.51  & H2 & \text{Second} & polar?
\\
7 & 267.86651 &	-29.58575 & 5144.97 & 0.99880 & 335 & 19.8 & 2.48 &-
& \text{Second} & IP?
\\
8 & 267.90982 &	-29.55845 & 5158.75 & 1.00000 & 121  & 30.6 & 2.30 &- & \text{Second} & polar?
\\
9 & 267.88075 &	-29.48562 & 5231.49 & 0.94949 & 347 & 188.4 & 1.65 &- &- & polar?
\\
10 & 267.91616 &	 -29.66900 & 5252.93 & 0.91425 & 211 & 134.8
	 &-&-&-& polar?
\\
11 & 267.83116 &	 -29.61651 & 5261.93 & 1.00000 & 438 & 28.0 & 1.46 & H10 & \text{Second} & IP?
\\
12 & 267.73141 &	 -29.49721 & 5334.76 & 0.99952 & 760 & 603.4 & 1.25 
	  &- & \text{Second} & polar?
\\
13 & 267.96901 &	 -29.58142 & 5501.16 & 0.99094 & 512 & 124.9 & 2.89
	 &- &- & polar?
\\
14$^\dag$ & 267.77173 & -29.61332 & 5608.21 & 0.96821 & 202  & 95.5 & 1.92 
 &-& \text{Third} & IP
\\
15 & 267.89161 &	 -29.46508 & 6335.85 & 1.00000 & 823 & 322.6 & 2.09  & H5  & \text{Second} & polar?
\\
16 & 267.89024 &	 -29.55369 & 6597.55 & 1.00000 & 487 & 24.3 & 4.30  & H9  & \text{Second} & polar?
\\ 
17 & 267.87162 &	 -29.49011 & 7448.98 & 0.99999  & 535 & 157.3 & 1.76  & H3 & \text{Second} & polar?
\\
18 & 267.78806 &	 -29.58177 & 7756.19 & 0.99941 & 214 & 54.5 & 4.76 &- &\text{Second} & IP?
\\
19 & 267.88203 &	 -29.49922 & 8546.28 & 1.00000  & 3402  & 132.7 & 3.27 &H4  & \text{Second} & IP?
\\
20 & 267.82785 &	 -29.50770 & 8844.82 & 0.90987 & 263 & 89.9 & 1.82
	&-  & \text{Second} & IP?
\\
21$^\ddag$ & 267.96375 & -29.55290 & 9877.52 & 0.99992  & 1963 & 153.1 & 1.44 & - &- & IP
\\
22$^\ddag$ & 267.96375 & -29.55290 & 10342.30 & 1.00000 & 1963 & 153.1 & 1.44 & H1  &- & IP 
\\
23 & 267.84487 &	 -29.57680 & 12002.70 & 1.00000 & 307 & 13.8 & 1.86 & H7 & \text{Second} & polar?
\\
24 & 267.90974 &	-29.71112 & 42219.03 & 1.00000  &1039 &343.0 &23.6  &- &- & DN?
\\
25 & 267.83142 &	 -29.65992 & 47317.12 & 0.98850 & 138 &77.0  &- &- &- & AB?
\\
\hline
\end{tabular}
\begin{tablenotes}
      \small
      \item 
      Notes:
      (1) Source sequence number assigned in the order of increasing period. The same source with dual periods is marked by \dag\ and \ddag. 
(2) and (3) Right Ascension and Declination (J2000) of the source centroid. 
(4) The modulation period determined by the GL algorithm.
(5) The probability of the periodic signal defined by Eqn.~\ref{A20}.  
(6) The number of total counts in the 1-8 keV band.
(7) The number of estimated background counts.
(8) The long-term variability index, defined as $\rm VI=S_{max}/S_{min}$, where $\rm S_{max}$ and $\rm S_{min}$ are the maximum and minimum photon fluxes among all the valid detections. Sources \#10 and \#25 have no measurable VI.
(9) The ID of previously detected periodic signals as given in table 2 of \cite{2012ApJ...746..165H}.
(10) Significant harmonics, if present.
(11) Tentative source classification.
\end{tablenotes}
\end{threeparttable}
\end{table*}

\section{X-ray spectral analysis}\label{sec:spectra}
We perform spectral analysis for all 23 confirmed periodic sources to gain insight on their nature. Source and background spectra are extracted from the same regions as described in Section~\ref{subsec:detect}, along with the ancillary response files (ARFs) and redistribution matrix files (RMFs), by using the CIAO tool \emph{specextract}. 
The spectra from individual observations are then coadded to form a combined spectrum of a given source, with the corresponding ARFs and RMFs weighted by the effective exposure. 
Further, the spectrum is adaptively binned to achieve a minimum of 20 counts and a signal-to-noise ratio (S/N) greater than 2 per bin.
The resultant spectra are analyzed using XSPEC v12.9.1.

Since these sources are expected to be CVs, we adopt a fiducial phenomenological spectral model, which consists of a bremsstrahlung continuum and three Fe lines \citep{2018ApJS..235...26Z,2019ApJ...882..164X}. 
We model each line with a Gaussian profile, fixing their line centroid at 6.40, 6.68 and 6.97 keV, respectively, and adopting a zero line width. 
All model components are subject to an unknown line-of-sight absorption (\emph{phabs} in XSPEC). 

It turns out that the plasma temperature ($T_{\rm b}$) is not well constrained in most sources, due to a moderate spectral S/N and the insufficient sensitivity of {\it Chandra} at energies above 8 keV. Hence for such cases we fix $T_{\rm b}$ at 40 keV, which is typical of IPs when their hard X-ray (up to tens of keV) spectra are available \citep{2016ApJ...818..136X,2016ApJ...826..160H}.
Setting a lower value of $T_{\rm b}$, for instance, at 20 keV, does not affect our following conclusions.

Physically, the three lines (hereafter referred to as the 6.4, 6.7 and 7.0 keV lines) correspond to Fe\,I K$\alpha$, Fe\,XXV K$\alpha$ and Fe\,XXVI Ly$\alpha$, respectively, which are among the most commonly detected emission lines in CV spectra. The latter two lines, in particular, arise from the post-shock plasma near the WD surface, and their flux ratio ($I_{7.0}/I_{6.7}$) has proven to be a robust tracer of the plasma temperature, hence also a good indicator of the WD mass \citep{1997ApJ...474..774F,1999ApJS..120..277E,2016ApJ...818..136X}.
Unfortunately, due to the limited number of counts, the Fe lines are insignificant in most of the 23 sources.
Only one source, \#21 (same as \#22), shows a significant ($\geq 3 \sigma$) line at all three energies. The 7.0 and 6.7 keV lines have a flux ratio of $I_{7.0}/I_{6.7} = 0.91^{+1.22}_{-0.56}$ (90\% errors), but the uncertainty is too large for a meaningful constraint on the WD mass. Three additional sources (\#2, \#20 and \#23) show a significant 6.7 keV only.

Results of the spectral analysis are summarized in Table \ref{tab:spec}. We have also derived the 1--8 keV unabsorbed luminosity based on the best-fit model and assuming a uniform distance of 8 kpc \citep{2009ApJ...705.1548R}.
It is noteworthy that source \#24 has an absorption column density of $4^{+2}_{-1}\times10^{21}{\rm~cm^{-2}}$, significantly lower than the expected column density of the LW, $N_{\rm H} \geq 7\times10^{21}{\rm~cm^{-2}}$ \citep{2011MNRAS.414..495R}. Hence 
this source is quite likely located in the foreground, and its true luminosity can be substantially lower than that listed in Table \ref{tab:spec}. 
Moreover, \#24 is the only source that shows a large flux variation, with VI = 23.6 (Table~\ref{tab:src}). A close examination of its long-term light curve indicates that it experienced an outburst caught by ObsID 9892 and 9893, two observations separated by 1 day. Therefore we have analyzed two additional spectra of \#24, one extracted from ObsID 9892 and 9893 and the other from the remaining observations. 
The best-fit plasma temperature is $6_{-2}^{+8}$ keV and $4_{-2}^{+10}$ keV in the outburst and quiescent state, respectively. 
The outburst has a 1--8 keV unabsorbed luminosity of $56^{+7}_{-7}\times10^{31}{\rm~erg~s^{-1}}$, compared to $3.6^{+0.8}_{-0.8}\times10^{31}{\rm~erg~s^{-1}}$ in the quiescent state, again for a distance of 8 kpc.

\begin{table*}
\centering
\begin{threeparttable}
\caption{X-ray spectral properties of the periodic sources \label{tab:spec}}
\begin{spacing}{1.19}

\begin{tabular}{lcccccc}
\hline
\hline
ID & $N_{\rm H}$ & $T_{\rm b}$ &  ${\rm EW}_{6.7}$ &$I_{6.7}$ & $\chi^2$/d.o.f & $L_{\rm 1-8}$ 
\\
(1) & (2) & (3) & (4) & (5) & (6) & (7)
\\
LW & $10^{22}\rm~cm^{-2}$ & keV & keV & $\rm 10^{-7}~ph~ cm^{-2}~s^{-1} $ & & $10^{31}\rm~erg~s^{-1}$ 
\\
\hline

1$^\dag$ & $0.5^{+0.8}_{-0.3}$ & 40 (fixed)  & - & - & 0.72/8  & $2.0^{+0.7}_{-0.6}$
\\
2 & $2.2^{+0.5}_{-0.5}$ & $21^{+57}_{-10}$ & $0.6^{+1.2}_{-0.3}$ & $2^{+1}_{-1}$ & 0.86/78  & $26^{+3}_{-2}$
\\
3 & $1.4^{+0.4}_{-0.3}$ & 40 (fixed) & - & - & 0.98/35  & $7^{+1}_{-1}$
\\
4 & $1.2^{+0.7}_{-0.5}$ & $11^{+53}_{-5}$ & - & - & 0.91/27  & $25^{+3}_{-5}$
\\
5 & $1.0^{+0.3}_{-0.2}$ & 40 (fixed) & - & -& 1.00/35 &  $6.2^{+0.9}_{-0.8}$
\\
6 & $1.8^{+0.6}_{-0.5}$ & $7^{+12}_{-3}$ 
	& - & -& 1.12/42 &  $10^{+1}_{-1}$
\\
7 & $1.6^{+0.6}_{-0.5}$ & $42^{+43}_{-32}$ & - &- & 0.86/39 &  $7 ^{+1}_{-1}$
\\
8 & $2^{+2}_{-1}$ & 40 (fixed) & - &- & 0.93/7 &  $2.0^{+0.6}_{-0.5}$
\\
9 & $1^{+1}_{-1}$ & $21^{+96}_{-20}$ & -  &- & 1.10/7 &  $4^{+1}_{-1}$
\\
10 & $3^{+7}_{-2}$ & 40 (fixed)  & - &- & 0.85/5 &  $2^{+2}_{-1}$
\\
11 & $1.9^{+0.4}_{-0.3}$ & 40 (fixed)  & - &- & 1.22/47 &  $11^{+1}_{-1}$
\\
12 & $6^{+10}_{-4}$ & 40 (fixed) &-&-& 0.78/6 &  $8^{+6}_{-3}$
\\
13 & $0.9^{+0.3}_{-0.2}$ & 40 (fixed) & - &- & 1.15/38 &  $9^{+1}_{-1}$
\\
14$^\dag$ & $0.5^{+0.8}_{-0.3}$ & 40 (fixed) & - &- &  0.72/8 &  $2.0^{+0.7}_{-0.6}$
\\
15 & $1.5^{+0.5}_{-0.4}$ & 40 (fixed) & - &- &  1.16/42 & $14^{+2}_{-2}$
\\
16 & $0.9^{+0.2}_{-0.2}$ & 40 (fixed) & - &-&  1.43/51  & $9.4^{+0.9}_{-0.9}$
\\
17 & $2.7^{+0.6}_{-0.5}$ & 40 (fixed)  & - &-&  0.75/36 & $13^{+2}_{-2}$
\\
18 & $2.0^{+0.9}_{-0.7}$ & 40 (fixed)  & -&- &  0.53/14  & $5^{+1}_{-1}$
\\
19 & $1.8^{+0.1}_{-0.1}$ & 40 (fixed)  & - &- &  0.95/185  & $93^{+4}_{-4}$
\\
20 & $2^{+2}_{-1}$ & $4^{+28}_{-3}$ & - &$1.4^{+0.9}_{-0.8}$ & 1.04/12  & $6^{+2}_{-2}$
\\
21$^\ddag$ & $2.8^{+0.3}_{-0.2}$ & 40 (fixed)  & $0.6^{+0.2}_{-0.2}$& 
$3^{+2}_{-1}$ & 1.10/147  & $67^{+4}_{-3}$
\\
22$^\ddag$ & $2.8^{+0.3}_{-0.2}$ & 40 (fixed)  &  $0.6^{+0.2}_{-0.2}$ & $3^{+2}_{-1}$ & 1.10/147  & $67^{+4}_{-3}$
\\
23 & $1.8^{+0.4}_{-0.4}$ & 40 (fixed)  & $1.2^{+0.7}_{-0.6}$  & $1.3^{+0.8}_{-0.6}$ & 1.43/37  & $10^{+1}_{-1}$
\\
24 & $0.4^{+0.2}_{-0.1}$ & $5^{+4}_{-2}$ & - &-&  1.19/55  & 
$10^{+1}_{-1}$
\\
25 & $1^{+3}_{-1}$ & $2^{+2}_{-1}$ &-&-&  0.80/2 & $0.8^{+0.3}_{-0.3}$ 
\\
\hline
\end{tabular}
\end{spacing}
\begin{tablenotes}
      \small
      \item
      Notes: 
      (1) Source sequence number as in Table~\ref{tab:src}. {\dag}\#1 and \#14 are the same source with two different periods; {\ddag}\#21 and \#22 are the same source with two different periods. 
(2) Line-of-sight absorption column density.
(3) The bremsstrahlung temperature. Fixed at a value of 40 keV if the spectrum provides no significant constraint to this parameter.
 (4) Equivalent width of the 6.7 keV line.
 (5) Integrated flux of the 6.7 keV line.
 (6) $\chi^2$ and degree of freedom of the best-fit model.
 (7) 1--8 keV unabsorbed luminosity for a distance of 8 kpc, corrected for the enclosed-energy fraction. Quoted errors are at the 90\% confidence level.
\end{tablenotes} 
\end{threeparttable}
\end{table*}

\section{Discussion}\label{sec:discussion}
In this section, we discuss the most significant implications of our results. We first provide a comparison between our work and \cite{2012ApJ...746..165H} (Section~\ref{subsec:compare}). Then, we try to classify the periodic sources based on their temporal and spectral properties, demonstrating that the majority of these sources are most likely magnetic CVs (Section~\ref{subsec:class}). Lastly, we attempt to constrain the fraction of various CV sub-classes in the Galactic bulge (Section~\ref{subsec:population}).

\subsection{Comparison with previous work} \label{subsec:compare}
\begin{figure*}
\centering
\includegraphics[width=0.49\textwidth]{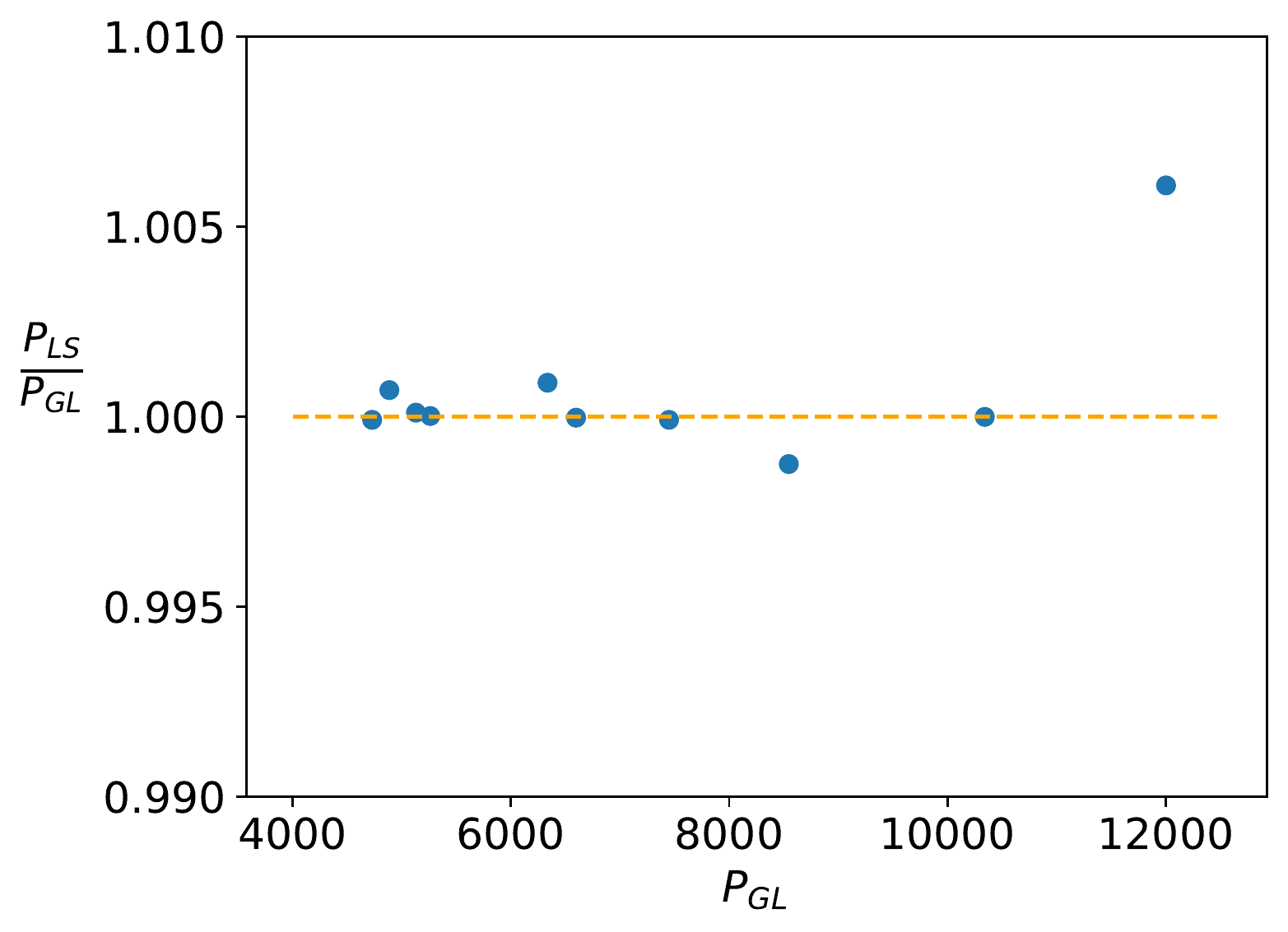}
\includegraphics[width=0.46\textwidth]{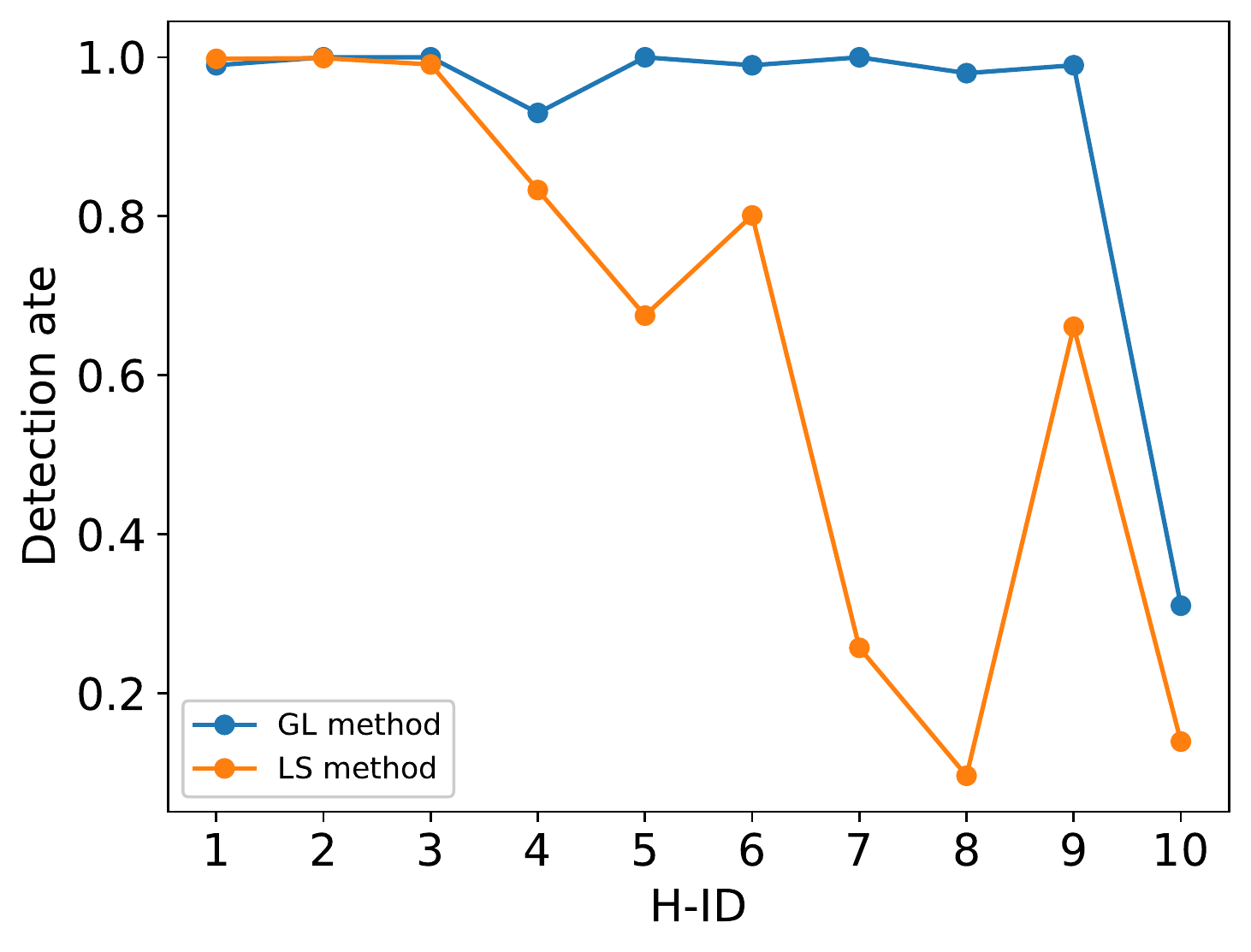}
\caption{Comparison between the period ({\it left}) and detection rate ({\it right}) of the 10 signals commonly found by the GL algorithm (this work) and the LS periodogram \citep{2012ApJ...746..165H}.The H-ID is the sequence number given in \citet{2012ApJ...746..165H}, same as listed in column 6 of Table \ref{tab:src}. 
\label{fig:com}}
\end{figure*}

We have detected 25 periodic signals from 23 sources in the LW (Table~\ref{tab:src}). 
Our detections fully recover the 10 periodic signals from 10 sources found by \cite{2012ApJ...746..165H}. 
As illustrated in the left panel of Figure~\ref{fig:com}, the measured periods of these 10 signals from the two studies agree with each other to within 0.7\% and show no systematic bias.
The remaining 15 periods are new detections. 
Since our work and \cite{2012ApJ...746..165H} have used the same set of {\it Chandra} observations, 
this difference must be owing to the different period searching methods employed.

\cite{2012ApJ...746..165H} employed the LS periodogram,
which, as described in Section~\ref{subsec:GL}, handles the photometric light curve, while the GL algorithm processes the phase-folded light curve with tolerance for observation gaps. We have applied the LS periodogram to our data in essentially the same way as \cite{2012ApJ...746..165H} and confirmed that only those 10 periods found in their work can be detected by this method. 
The other 15 periods do not result in a significant detection in the LS periodogram, mainly due to its low efficiency with low-count sources.
The detection rate 
never exceeds 20\% for net counts $\lesssim$ 150, according to the simulations presented in figure 7 of \citet{2012ApJ...746..165H}.

For comparison, we provide an estimate on the detection rate of the 10 periodic signals that are detected in this work as well as by \citet{2012ApJ...746..165H}, when using the GL algorithm. For practical purposes we assume an intrinsic sinusoidal shape (Eqn.~\ref{eqn:sin}), for which the total counts and period are taken directly from the observed values, whereas the relative amplitude ($A_0$) is taken 
from \citet{2012ApJ...746..165H}, which was based on the Rayleigh statistics \citep{1983A&A...128..245B,2003ApJ...599..465M}. 
For each of the 10 signals, 100 simulated light curves are produced and fed to the GL algorithm.
Again, we take the 90\% probability threshold and a period accuracy of 1\% to define a valid detection. The percentage of valid detections 
is plotted in the right panel of Figure~\ref{fig:com}, along with the detection rate of the LS periodogram taken from \citet{2012ApJ...746..165H}.
Clearly, the GL algorithm works better in almost all cases compared to the LS periodogram. 


\begin{figure*}
\centering
\includegraphics[scale=0.75]{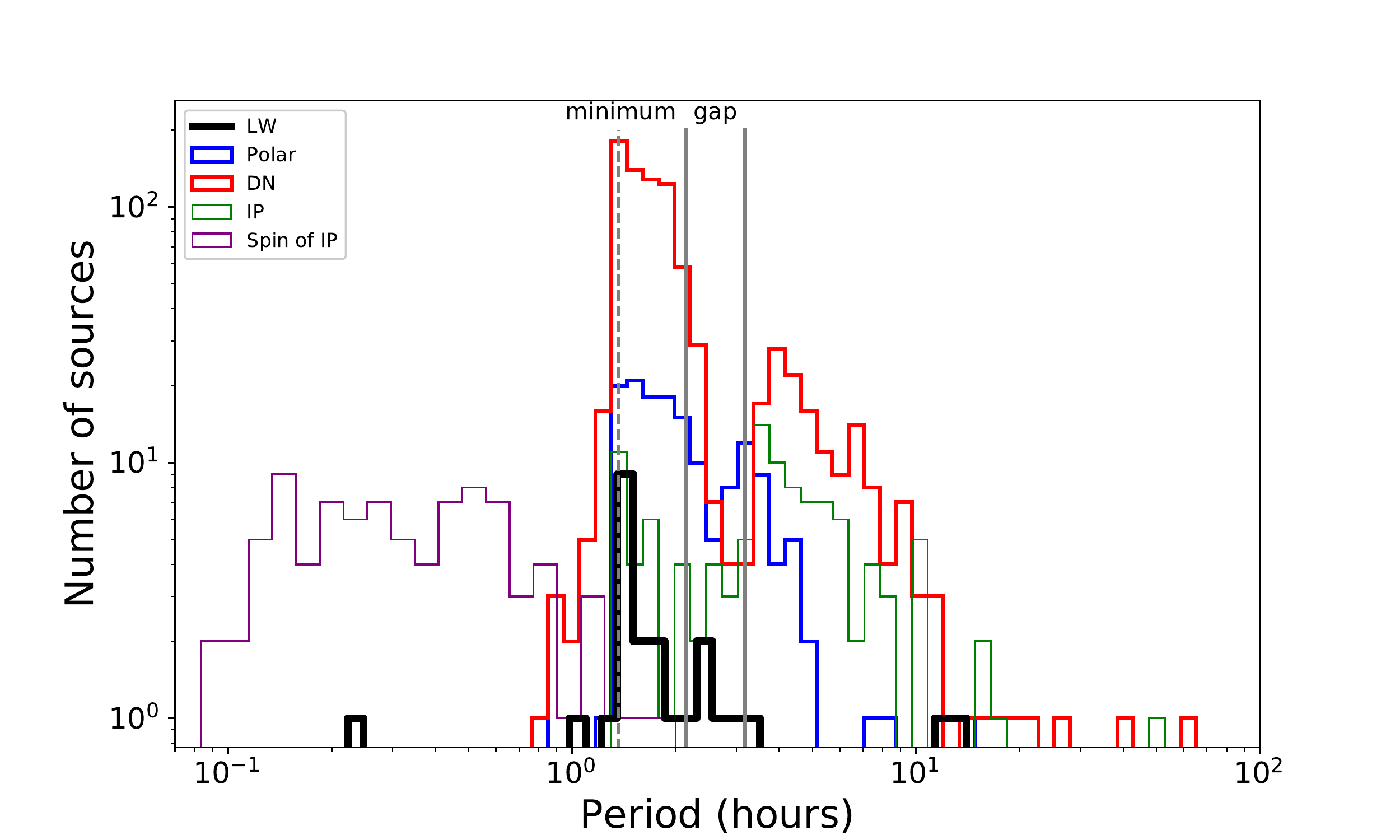}
\caption{The period distribution of the LW sources (black histogram), in comparison with the spin (purple) and orbital (green) periods of IPs, the orbital period of polars (blue), and the orbital period of DNe from the catalog of \citet{2003A&A...404..301R}, version 7.20. 
Note that a cut of 300 sec is applied for the spin period, to be consistent with the lower bound of our period searching.
The period gap of CVs is delineated by a pair of vertical solid lines, and the period minimum is marked by a vertical dashed line, the values of which are taken from \citet{2011ApJS..194...28K}.\label{fig:N_P}}
\end{figure*}

\subsection{Classifying the periodic X-ray sources}
\label{subsec:class}
While it is generally difficult to unambiguously identify the nature of an X-ray source without knowing its optical counterpart, which is the case for most LW sources, the X-ray temporal and spectral properties of the 23 periodic sources contain useful information that actually allow for a reasonable classification for many of them. 

First of all, all but one (\#25) of these sources are found in the 1--8 keV luminosity range of $10^{31}-10^{33}\rm~erg~s^{-1}$, assuming a distance of 8 kpc (Table~\ref{tab:spec}), which is typical of CVs. 
On the contrary, coronally active binaries (ABs), which are thought to have a substantial contribution to the detected X-ray sources in the LW \citep{2009Natur.458.1142R}, generally have unabsorbed X-ray luminosities below $\rm 10^{31}~erg~s^{-1}$ \citep{2006A&A...450..117S}. 
This stems from the empirical fact that the coronal X-ray emission from low-mass stars saturates at $\sim$10$^{-3}$ of the bolometric luminosity \citep{2004A&ARv..12...71G}.
The X-ray spectra of these periodic sources are less informative, since except in a few cases they lack the unambiguous sign of Fe lines due to the moderate S/N (Section~\ref{sec:spectra}), which otherwise would be another characteristic signature of CVs (e.g., \citealp{2016ApJ...818..136X}).
Nevertheless, most of these sources do show a hard continuum that is again typical of CVs.
Furthermore, the vast majority of the detected periods fall between 1.1--3.3 hours, consistent with the range of short-period CVs in the solar neighborhood \citep{2003A&A...404..301R}, i.e., those having an orbital period between the so-called period minimum and the upper bound of the period gap, as illustrated in Figure~\ref{fig:N_P}. 
Therefore, the global X-ray properties point to CVs dominating the detected periodic sources, a conclusion also drawn by \cite{2012ApJ...746..165H}.

It will then be interesting to ask to which sub-class of CVs these periodic sources belong. The phase-folded light curves may provide useful hints to this question. 
In magnetic CVs, including polars and IPs, their spin modulations can give rise to a characteristic light curve. For polars, let us consider the simplistic situation in which hard X-rays (photon energy $\gtrsim$1 keV) are produced in only one of the two magnetic poles\footnote{In this picture, accretion can still take place in the other pole at an even much higher accretion rate, producing quasi-black-body emission peaking at the soft X-ray and extreme ultraviolet bands \citep{2001cvs..book.....H}.}.
Denoting $i$ the angle between the line-of-sight and the spin axis, $\beta$ the angle between the spin and magnetic axes, and $\epsilon$ the magnetic colatitude of the accretion column, when $i+\beta+\epsilon > 90^{\circ}$, the two poles will alternately drift across the front side of the WD. 
The resultant light curve will look like that of source \#6 , \#8 and \#10, with nearly half of the cycle showing a valley of near-zero hard X-ray flux \citep{1985A&A...148L..14H}. 
The periods of these sources are consistent with the period range of known polars (for polars the orbital and spin periods are equal; Figure~\ref{fig:N_P}).
This so-called ``two-pole'' behavior becomes more complicated if some hard X-rays were also produced near the second pole, in which case the ``valley'' is partially filled, resulting in a light curve like that of \#4, \#5, \#9, \#12, \#13, \#15 ,\#16, \#17, and \#23. 
On the other hand, if $i+\beta+\epsilon < 90^{\circ}$, the hard X-ray-emitting pole would be always visible (the so-called ``one-pole'' behavior), producing a roughly constant light curve, although under certain condition ($\beta < i$) dips can be present due to obscuration by the accretion stream \citep{2001cvs..book.....H}. The ``one-pole'' case generally does not favor a robust detection of the period. 

IPs share the above two-pole and one-pole behavior. 
In addition, orbital modulation can result in obscuration by the companion star,
by the accretion stream, or by the ``disc overflow'' in the presence of an truncated accretion disk \citep{1996MNRAS.280..937N}. 
Thus the light curve shape of IPs is often complex and can exhibit both sinusoidal variations and dips. Examples of this kind are found in sources \#3, \#7, \#11, \#18 and \#20, although we cannot rule out the possibility of ``one-pole'' polars obscured by the accretion stream.
We also consider source \#19 a likely IP, for it has the highest X-ray luminosity ($9.3\times10^{32}{\rm~erg~s^{-1}}$) among all sources. Notably, its sinusoidal-like variation also has the lowest amplitude ($\sim$20\%) among all sources, which might be due to orbital modulation by a disc overflow. 

The most robust identification of IPs is to detect both the spin and orbital periods. Among our sources, \#1/\#14 and \#21/\#22 show dual periods.
For \#1, the period is 853.8 sec and the corresponding phase-folded light curve is consistent with the two-pole behavior (Figures~\ref{fig:pCV_sample_1}). This is naturally understood in terms of a spin modulation. 
In the meantime, the phase-folded light curve of \#14 shows a dip at phase 0.9--0.1, which may be understood as obscuration by the accretion stream, i.e., an orbital modulation. The corresponding orbital period of 5608.2 sec is reasonable for an IP. 
The two periods of \#21 and \#22 differ by only 5\% from each other (9877.5 sec vs. 10342.3 sec).
The light curve of \#22 exhibits a prominent narrow dip near phase 0.5 (Figure~\ref{fig:pCV_sample_2}). This is a clear sign of eclipse when viewed from a high inclination, thus the period of \#22 should be the orbital period. 
On the other hand, the light curve of \#21 resembles the two-pole behavior and is best understood as due to spin modulation. 
In this regard, source \#21/\#22 is probably an IP or a so-called asynchronous polar. 

It is worth asking, if some of the aforementioned sources (e.g., \#3, \#7, \#11, \#18, \#19, \#20) were indeed IPs, why only their orbital period is detected. One plausible explanation is that their spin period, presumably below $\sim$1 hour, escapes detection due to a relatively low detection efficiency. Our simulations (Section \ref{subsec:simulation}) suggest that for moderate total counts and moderate variation amplitude, the detection rate is generally low for short periods. This effect may apply in all those six candidate IPs except \#19, which has total counts of 3402. Perhaps this source has a very low spin modulation due to the ``one-pole'' behavior.
On the other hand, source \#2, which has not been discussed so far, shows a period of 3820.8 sec, significantly lower than the canonical orbital period minimum of $\sim$82 minutes determined from both observations \citep{2009MNRAS.397.2170G} and theoretical modelling \citep{2011ApJS..194...28K}. 
Thus this period is more likely the spin period.
In this regard, a spike-like feature in the phase-folded light curve of \#2 may be caused by a bright spot on the WD surface. 
The unseen orbital modulation may be due to a low inclination angle. 

In DNe (i.e., non-magnetic CVs), hard X-rays are produced in the boundary layer near the WD surface, in which case spin modulation is essentially absent and orbital modulation is also expected to be weak, unless the inclination angle is sufficiently large to allow for a total eclipse. Among the 25 periodic signals, no clear sign of total eclipse is seen except for the case of \#22. However, we have argued in the above that source \#21/\#22 is an IP given the dual periods. Therefore, we probably have not detected any eclipsing DN among the periodic sources. This is consistent with the low detection rates for eclipse predicted by our simulations in Section~\ref{subsec:simulation}. 

The nature of source \#24, which has the second longest period (42219.0 sec), deserves special attention. At a glance, it has a sinusoidal-like light curve similar to that of \#19, which is indicative of an IP. However, the spectrum of \#24, characterized by a plasma temperature of $5^{+4}_{-1}$ keV, is both too soft for an IP \citep{2016ApJ...818..136X}
and too high for an AB \citep{2004A&ARv..12...71G}. 
While this temperature is not atypical of polars, the period of \#24 is much larger than that of any previous known polars (Figure~\ref{fig:N_P}). 
This leads us to consider \#24 a DN, which is consistent with it having experienced an outburst.  
In this case, its low-amplitude ($\sim$35\%) light curve may be due to a broad absorption dip (phase 0.8-0.3), similar to that seen in the X-ray light curve of the famous DN, Z Chamaleontis \citep{2011A&A...536A..75N}. Such a dip was suggested to be caused by the absorption of dense gas clouds as the result of disk overflow. 
Therefore, we seem to have caught one DN among the periodic sources, although this source is quite likely located in the foreground (Section~\ref{sec:spectra}).

Lastly, we remark on source \#25, for which we have found no satisfying classification. This source has an unabsorbed 1--8 keV luminosity of $\rm 0.8\times 10^{31}~erg~s^{-1}$, the lowest among all 23 sources. Its X-ray spectrum, characterized by a best-fit plasma temperature of 2 keV (notably with large uncertainties; Table~\ref{tab:spec}), is also the softest among all sources.
Together these values argue against a CV, but are more typical of ABs. 
Moreover, the phase-folded light curve of \#25 suggests that the X-ray flux drops to the background level for nearly half of the 47317-sec period (Figure~\ref{fig:Figure_p}). 
As mentioned in the above, this may be due to a polar producing hard X-rays from only one pole.
However, we consider such a case rather unlikely, since the corresponding period again greatly exceeds the empirical range of known polars.
Alternatively, the light curve can be explained by a total eclipse lasting for nearly half cycle, 
suggesting an AB system viewed nearly edge-on.
However, this would require the obscuring star to have a size not much smaller than the orbital separation and in the meantime negligible X-ray emission.
One possibility is that this system consists of a Sun-like star and an A-star, the latter producing no significant X-rays due to a very weak surface magnetic field \citep{2004A&ARv..12...71G}. 
An orbital separation of 4 $R_\odot$ estimated from the period is compatible with such a system. 

In summary, based on the X-ray properties of the periodic sources, in particular their phase-folded light curves, we identify 12 candidate polars, 2 probable plus 7 candidate IPs, one likely DN and one possible AB. 
Admittedly, polars and IPs share similar characteristics in their phase-folded light curves, as described above. 
Hence some of the polar candidates may turn out to be IPs, and vice versa.
Moreover, the non-detection of the spin period (or orbital period in the case of \#2) brings about substantial uncertainty in the IP identification.

It is also noteworthy that only two long ($>3.5$ hours) periods are found among the 22 orbital periods (3 probable spin periods subtracted), and in fact neither of them are genuine CVs in the LW (\#24 is a foreground source and \#25 is likely an AB).
On the contrary, long-period magnetic CVs (IPs plus polars) occupy a fraction of $\sim$30\% in all magnetic CVs in the solar neighborhood, according to the catalog of \citet{2003A&A...404..301R}, version 7.20, although this catalog is likely biased toward luminous (thus long-period) CVs.
A recent census of CVs within 150 pc from the Sun based on {\it Gaia} distances also finds that long-period CVs account for a fraction of 17\% \citep{2020MNRAS.494.3799P}.
The lack of long-period CVs in the LW cannot be due to a selection effect of the GL algorithm, because our simulations in Section~\ref{subsec:simulation} predict that the detection rate is generally higher for longer periods.
\cite{2012ApJ...746..165H} noticed the narrow period distribution (1--3 hours) of the 10 sources they detected and the similarity with the period distribution of polars in the solar neighborhood (Figure~\ref{fig:N_P}), on the basis of which they suggested that all these 10 sources are polars. 
We now provide strong evidence that at least one of these 10 sources is an IP, given its dual periods (\#21/\#22).
We suggest that the absence of long-period CVs could be due to an age effect, in the sense that CVs in the LW (inner Galactic bulge) are predominantly old and have substantially shrunk their orbits, while CVs in the solar neighborhood can comprise of younger populations and have presently wider orbits.  

\subsection{CV populations in the inner Galactic bulge}\label{subsec:population}
In the previous section, we have classified, with a varied degree of confidence, the 23 periodic sources, finding that the vast majority of them are either polars or IPs. Based on this classification, we now take one step forward to place constraints on the fraction of different sub-classes of CVs in the inner Galactic bulge, provided that the LW is a representative field. 

The observed number of a given sub-class, for instance, polars, can be expressed as,
\begin{equation}
  N_{\rm det,polar} = P_{\rm det} \times g \times \alpha_{\rm polar} \times N_{\rm tot},
\end{equation}
where $N_{\rm tot}$ is the total number of CVs in the LW, $\alpha_{\rm polar}$ is the intrinsic fraction of polars among all CVs, $g$ is a geometric factor that determines the detectability of a periodic signal due to spin and/or orbital modulations, and $P_{\rm det}$ is the detection rate given the available data and the GL algorithm. Here we take $N_{\rm tot}$ = 600 ($=667\times90\%$), effectively subtracting 10\% of all LW sources with 1--8 keV total counts great than 100, which is roughly the fractional contribution from the cosmic X-ray background in the LW \citep{2018ApJS..235...26Z}. This implicitly assumes that all 600 sources are CVs and that CVs with similar or larger total counts have been detected in 100\% completeness. 

We further assume that the substantial sinusoidal variation occurs only in the ``two-pole'' situation, that is, only one pole produces the hard X-rays and for this pole $i+\beta+\epsilon > 90^{\circ}$. 
Adopting an 50\% probability to fulfill each of these two conditions, this allows us to take $g_{\rm polar}$ = $50\%\times50\% = 25\%$. 

\begin{figure}
\includegraphics[scale=0.45]{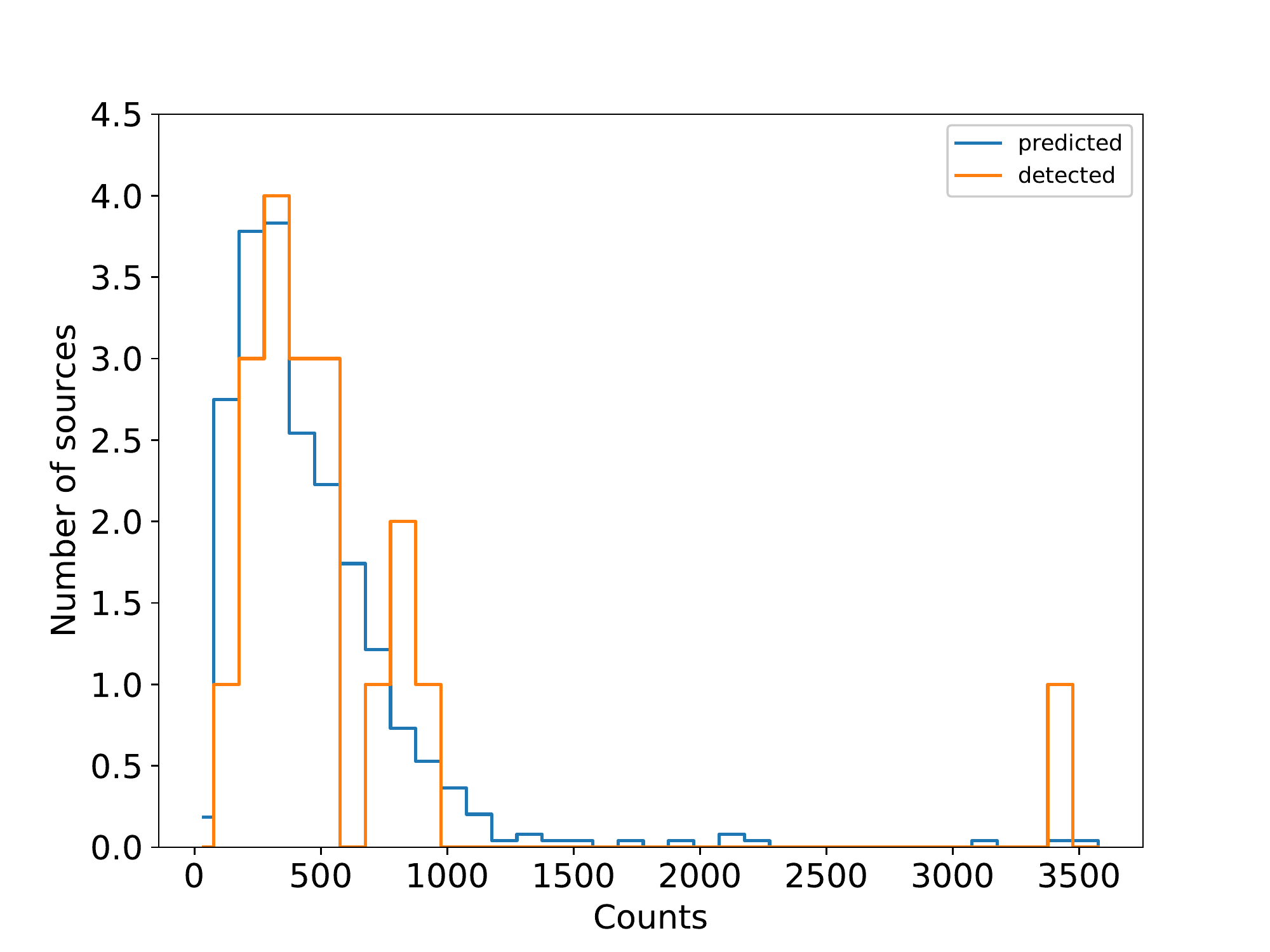}
\caption{Comparison between the number of predicted and actually detected polars, as a function of detected source counts. \label{fig:NP_sim}}
\end{figure}

In reality, $P_{\rm det}$ depends on the period, variation amplitude and total counts (Section~\ref{subsec:simulation}). 
Without a {\it  priori} information about period and amplitude, we make use of our simulation results to estimate $P_{\rm det}$ with the following approximations:
(i) The short and long orbital periods (below or above 3.2 hours) are represented by $P$=5540 sec and $P$=45540 sec, respectively. A weight of 82\% (18\%) is assigned to the short (long) period, roughly according to the period distribution of solar neighborhood polars; 
(ii) the amplitude is evenly sampled between 0.5 to 0.9, neglecting the contribution from any lower amplitudes (i.e., $P_{\rm det} = 0$ in this case);
(iii) the observed source counts are grouped into different bins, from $C$=75 to 3575 cts at a step of 100 cts.  
Then in each bin the number of polars is predicted to be,
\begin{equation}
N_{\rm det,polar}^i =  g_{\rm polar} \times \alpha_{\rm polar} \times P_{\rm det}^i \times N^i,
\end{equation}
where $N^i$ is the intrinsic number of sources in the $i$th bin, satisfying $\sum{N^i}=N_{\rm tot}=600$.  
$N_{\rm det,polar}^i$ is constrained by the number of actually detected polars in a given bin.
We take $\sum{N_{\rm det,polar}^i} = 19$, effectively counting all the periodic sources except the two probable IPs, one possible AB and one foreground DN (Section~\ref{subsec:class}).
Figure~\ref{fig:NP_sim} compares the number of detected and predicted polars as a function of total counts, which apparently agree well with each other.  
The required normalization of the predicted numbers is such that $\alpha_{\rm polar} \approx 18\%$. Strictly speaking, this value is an upper limit, since a fraction of the 19 periodic sources could be IPs and $g_{\rm polar}$ may have been underestimated. 
For comparison, the fraction of polars is 13.5\% among all known CVs in the solar neighborhood  \citep{2003A&A...404..301R}. This lends support to our above working assumptions.

A similar approach can be applied to constrain the fraction of IPs, by evaluating the detectability of spin modulation. 
We still adopt $g_{\rm IP} = 25\%$ for detectable two-pole behavior. The detection rate of the GL algorithm is now represented by our simulations having a test period of $P=554$ sec, which is suitable for the spin period. In Section~\ref{subsec:class}, we identify three periods as probable spin period of IPs (\#1, \#2 and \#21). 
Hence, according to their total counts, we estimate the fraction of IPs to be $\alpha_{\rm IP} \approx 5\%$ in the LW. 
This is to be contrasted with 8.6\%, the observed fraction of IPs in the solar neighborhood \citep{2003A&A...404..301R}.

The above estimates thus imply that $\lesssim$23\% of all CVs in the LW are magnetic (polars plus IPs), although we shall note that in our approach the assumed geometry for polars and IPs are not mutually complementary. 
By face value, the fraction of magnetic CVs in the LW is similar to that in the solar neighborhood ($\sim$23\%). However, we shall caution that both the LW sources and the  \citet{2003A&A...404..301R} catalog are subject to substantial selection bias. The recent volume-limited sample of CVs within 150 pc from the Sun \citep{2020MNRAS.494.3799P} suggests a remarkably higher fraction of 36\% for magnetic CVs, although the size of this sample is still small (42 in total). 

\begin{figure}
\centering
\includegraphics[scale=0.45]{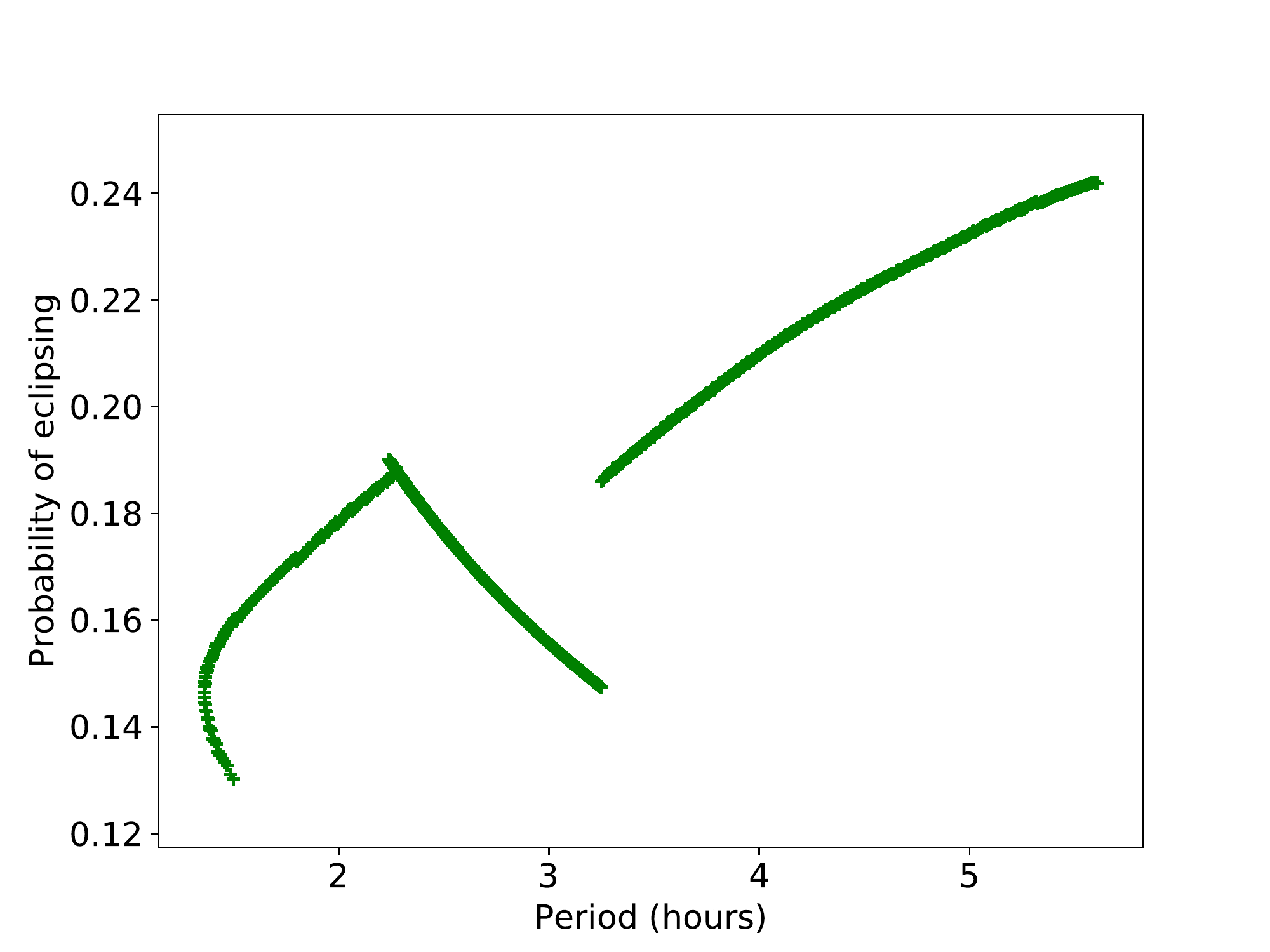}
\caption{Probability of eclipsing CVs. The break around 3.2 hours results from different mechanisms of angular momentum loss above and below the period gap \citep{2011ApJS..194...28K}.
The overall descending trend with decreasing periods is due to the faster shrinking of the donor radius than the shrinking of the orbit.
The reverse trend within the period gap 
results from the system evolving as a detached binary with a decreasing period and nearly unchanged donor radius.
\label{fig:simpCV}}
\end{figure}

We now turn to the fraction of non-magnetic CVs. As discussed in Section~\ref{subsec:class}, at most one DN is found among the periodic sources. Due to the general absence of spin modulation and weak orbital modulation on the X-ray light curve, the most likely cause of a prominent periodic signature in a DN should be a total eclipse. 
The geometry of a WD eclipse is relatively well defined, given the inclination angle ($i$), the WD radius ($R_{\rm WD}$), the radius of the donor star ($R_2$) and the orbital separation ($a$), together satisfying: 
\begin{equation}
{a \times {\rm cos}i}\leq {R_{\rm WD}+R_2}.
\end{equation}
Physically motivated values of $R_2$ and $a$ can be obtained from synthesis simulations of CVs \citep{2011ApJS..194...28K}, while $R_{\rm WD}$ is fixed at a value corresponding to $M_{\rm WD} = 0.75 {\rm~M_\odot}$.
Assuming that the inclination angle is evenly distributed between $0^\circ$--$90^\circ$, the probability of eclipsing, $g_{\rm DN}$, is derived and plotted as a function of orbital period (related to $a$ via Kepler's third law) in Figure \ref{fig:simpCV}. 
$g_{\rm DN}$ ranges between $0.20-0.24$ for long-period CVs and 
$0.12-0.20$ for short-period CVs. 

We then utilize the simulation results for eclipse (Figure~\ref{fig:eclipse}) to estimate the detection probability using the GL algorithm. Specifically, we assign a weight of 83\% (17\%) for the representative period of 5258 (15258) sec for cases below (above) the period gap. These fractions are again taken from the statistics of solar neighborhood DNe \citep{2003A&A...404..301R}. 
A substantial uncertainty lies in the total counts. The X-ray luminosity of DNe are typically below $\rm 10^{32}~erg~s^{-1}$ \citep{2016ApJ...818..136X}, hence we consider only the detection rates for total counts below 575, very roughly equaling to this luminosity threshold.
Requiring that the number of detectable eclipsing DNe is no larger than one,  
we thus place an upper limit of $\sim$62\% for $\alpha_{\rm DN}$, the intrinsic fraction of DNe in the LW. 

This value is apparently incompatible with our naive anticipation. It is expected that non-magnetic CVs comprise of $\sim$80\% of all CVs, if the percentage of $\sim$23\% for magnetic CVs estimated in the above is correct, or if the inner Galactic bulge has a similar fraction of non-magnetic CVs as in the solar neighborhood.  
We should caution that the estimate of $\alpha_{\rm DN}$ is rather sensitive to the presumed fraction of short/long periods, since the detection rate of eclipse depends sensitively on the period (Figure~\ref{fig:eclipse}). 
For instance, raising the weight of short-period DNe to 91\% will bring $\alpha_{\rm DN}$ to a value of 77\%.
A larger fraction of short-period DNe in the Galactic bulge than in the solar neighborhood is plausible, as pointed out in Section~\ref{subsec:class}, if bulge CVs are predominantly old. 
Synthesis simulations of CVs suggest that the evolution timescale above the period gap is only 0.24 Gyr from the onset of mass transfer \citep{2011ApJS..194...28K}.
We may reverse the above argument by emphasizing that no single long-period (above 4 hours) DN is detected in the inner bulge (precluding source \#24, which is located in the foreground), despite a moderate-to-high detection rate predicted for long-period eclipsing sources. 
Indeed, the presence of merely 20 long-period CVs with $C \approx 500$ in the inner bulge are sufficient to produce one detection of eclipsing source, according to the probability in Figure~\ref{fig:simpCV} and the detection rates in Figure~\ref{fig:eclipse}.  

\section{Summary}\label{sec:summary}
We have searched for periodic X-ray sources in the Limiting Window, utilizing $\sim$1 Ms {\it Chandra} observations and the Gregory-Loredo algorithm. We have also generated a large set of simulated light curves, based on idealized sinusoidal and piecewise functions, to evaluate the detection completeness of the GL algorithm. Our main findings include:

\begin{itemize}
\item We have detected 23 periodic sources with 25 signals, among a parent list of 667 discrete X-ray sources in the LW. Among them, 10 signals were previously found by \cite{2012ApJ...746..165H} using the Lomb-Scargle periodogram, whereas the remaining 15 signals are new discoveries. This underscores the advantage of the GL algorithm in dealing with relatively faint sources and irregular X-ray observations.

\item The vast majority of the 23 periodic sources are classified as magnetic CVs, based on their period range, X-ray luminosities, spectral properties and phase-folded light curves. In particular, dual periods are found in two sources, providing strong evidence that they are IPs. 

\item In addition, one source is classified as a DN, based on its long period, soft X-ray spectrum and an outburst caught by the {\it Chandra} observations. However, its low line-of-sight absorption indicates that it is most likely a foreground source. Another source, which has the longest period among all, could be an eclipsing AB, in view of it soft X-ray spectrum, low X-ray luminosity and phase-folded light curve characterized by a wide dip. 

\item Based on the number of tentatively classified periodic sources, a simplified geometry involved in different CV sub-classes, and the detection completeness evaluated from our simulations, we are able to provide meaningful constraints on the fraction of polars and IPs in the inner Galactic bulge, which is $\lesssim$18\% and 5\%, respectively. This suggests that the fraction of magnetic CVs (polars plus IPs) in the inner bulge is similar to that in the solar neighborhood. On the other hand, the lack of candidate periodic DNe in the inner bulge and the low detection completeness expected for eclipsing sources result in a large uncertainty in the estimated fraction of DNe. 
 
\item The apparent lack of long-period CVs in the LW suggests an intrinsically different period distribution, in the sense that CVs in the inner bulge are predominantly old and have substantially shrunk their orbits. 

\end{itemize}
It will be interesting to further compare the period distribution of the LW CVs with that of other stellar populations, especially in dense environments including globular clusters and the Galactic center. We will present such a study in future work, utilizing the techniques developed in this work.

\vskip0.5cm
\noindent{\bf Acknowledgements}\\
We are grateful to the late Mikhail G. Revnivtsev, a friend and colleague, whose insight led to the conduction of most {\it Chandra} observations used this work.
We thank Xiangdong Li, Xiaojie Xu and Zhenlin Zhu for helpful discussions. This work is supported by the National Key Research and Development Program of China under grant 2017YFA0402703.
\\

\noindent{\bf Data Availability}\\
The data underlying this article will be shared on reasonable request to the corresponding author. 

\bibliography{sample63}{}

\begin{thebibliography}{}
\makeatletter
\relax
\def\mn@urlcharsother{\let\do\@makeother \do\$\do\&\do\#\do\^\do\_\do\%\do\~}
\def\mn@doi{\begingroup\mn@urlcharsother \@ifnextchar [ {\mn@doi@}
  {\mn@doi@[]}}
\def\mn@doi@[#1]#2{\def\@tempa{#1}\ifx\@tempa\@empty \href
  {http://dx.doi.org/#2} {doi:#2}\else \href {http://dx.doi.org/#2} {#1}\fi
  \endgroup}
\def\mn@eprint#1#2{\mn@eprint@#1:#2::\@nil}
\def\mn@eprint@arXiv#1{\href {http://arxiv.org/abs/#1} {{\tt arXiv:#1}}}
\def\mn@eprint@dblp#1{\href {http://dblp.uni-trier.de/rec/bibtex/#1.xml}
  {dblp:#1}}
\def\mn@eprint@#1:#2:#3:#4\@nil{\def\@tempa {#1}\def\@tempb {#2}\def\@tempc
  {#3}\ifx \@tempc \@empty \let \@tempc \@tempb \let \@tempb \@tempa \fi \ifx
  \@tempb \@empty \def\@tempb {arXiv}\fi \@ifundefined
  {mn@eprint@\@tempb}{\@tempb:\@tempc}{\expandafter \expandafter \csname
  mn@eprint@\@tempb\endcsname \expandafter{\@tempc}}}

\bibitem[\protect\citeauthoryear{{Buccheri} et~al.,}{{Buccheri}
  et~al.}{1983}]{1983A&A...128..245B}
{Buccheri} R.,  et~al., 1983, \aap, \href
  {https://ui.adsabs.harvard.edu/abs/1983A&A...128..245B} {128, 245}

\bibitem[\protect\citeauthoryear{{Cicuttin}, {Colavita}, {Cerdeira}, {Mutihac}
  \& {Turrini}}{{Cicuttin} et~al.}{1998}]{1998ApJ...498..666C}
{Cicuttin} A.,  {Colavita} A.~A.,  {Cerdeira} A.,  {Mutihac} R.,   {Turrini}
  S.,  1998, \mn@doi [\apj] {10.1086/305564}, \href
  {https://ui.adsabs.harvard.edu/abs/1998ApJ...498..666C} {498, 666}

\bibitem[\protect\citeauthoryear{{Deeming}}{{Deeming}}{1975}]{1975Ap&SS..36..137D}
{Deeming} T.~J.,  1975, \mn@doi [\apss] {10.1007/BF00681947}, \href
  {https://ui.adsabs.harvard.edu/abs/1975Ap&SS..36..137D} {36, 137}

\bibitem[\protect\citeauthoryear{{Downes}, {Webbink}, {Shara}, {Ritter}, {Kolb}
   \& {Duerbeck}}{{Downes} et~al.}{2001}]{2001PASP..113..764D}
{Downes} R.~A.,  {Webbink} R.~F.,  {Shara} M.~M.,  {Ritter} H.,  {Kolb} U.,
  {Duerbeck} H.~W.,  2001, \mn@doi [\pasp] {10.1086/320802}, \href
  {https://ui.adsabs.harvard.edu/abs/2001PASP..113..764D} {113, 764}

\bibitem[\protect\citeauthoryear{{Edelson} \& {Krolik}}{{Edelson} \&
  {Krolik}}{1988}]{1988ApJ...333..646E}
{Edelson} R.~A.,  {Krolik} J.~H.,  1988, \mn@doi [\apj] {10.1086/166773}, \href
  {https://ui.adsabs.harvard.edu/abs/1988ApJ...333..646E} {333, 646}

\bibitem[\protect\citeauthoryear{{Ezuka} \& {Ishida}}{{Ezuka} \&
  {Ishida}}{1999}]{1999ApJS..120..277E}
{Ezuka} H.,  {Ishida} M.,  1999, \mn@doi [\apjs] {10.1086/313181}, \href
  {https://ui.adsabs.harvard.edu/abs/1999ApJS..120..277E} {120, 277}

\bibitem[\protect\citeauthoryear{{Fabbiano}}{{Fabbiano}}{2006}]{2006ARA&A..44..323F}
{Fabbiano} G.,  2006, \mn@doi [\araa] {10.1146/annurev.astro.44.051905.092519},
  \href {https://ui.adsabs.harvard.edu/abs/2006ARA&A..44..323F} {44, 323}

\bibitem[\protect\citeauthoryear{{Fujimoto} \& {Ishida}}{{Fujimoto} \&
  {Ishida}}{1997}]{1997ApJ...474..774F}
{Fujimoto} R.,  {Ishida} M.,  1997, \mn@doi [\apj] {10.1086/303483}, \href
  {https://ui.adsabs.harvard.edu/abs/1997ApJ...474..774F} {474, 774}

\bibitem[\protect\citeauthoryear{{G{\"a}nsicke} et~al.,}{{G{\"a}nsicke}
  et~al.}{2009}]{2009MNRAS.397.2170G}
{G{\"a}nsicke} B.~T.,  et~al., 2009, \mn@doi [\mnras]
  {10.1111/j.1365-2966.2009.15126.x}, \href
  {https://ui.adsabs.harvard.edu/abs/2009MNRAS.397.2170G} {397, 2170}

\bibitem[\protect\citeauthoryear{{Gregory} \& {Loredo}}{{Gregory} \&
  {Loredo}}{1992}]{1992ApJ...398..146G}
{Gregory} P.~C.,  {Loredo} T.~J.,  1992, \mn@doi [\apj] {10.1086/171844}, \href
  {https://ui.adsabs.harvard.edu/abs/1992ApJ...398..146G} {398, 146}

\bibitem[\protect\citeauthoryear{{G{\"u}del}}{{G{\"u}del}}{2004}]{2004A&ARv..12...71G}
{G{\"u}del} M.,  2004, \mn@doi [\aapr] {10.1007/s00159-004-0023-2}, \href
  {https://ui.adsabs.harvard.edu/abs/2004A&ARv..12...71G} {12, 71}

\bibitem[\protect\citeauthoryear{{Hailey} et~al.,}{{Hailey}
  et~al.}{2016}]{2016ApJ...826..160H}
{Hailey} C.~J.,  et~al., 2016, \mn@doi [\apj] {10.3847/0004-637X/826/2/160},
  \href {https://ui.adsabs.harvard.edu/abs/2016ApJ...826..160H} {826, 160}

\bibitem[\protect\citeauthoryear{{Heise}, {Brinkman}, {Gronenschild}, {Watson},
  {King}, {Stella}  \& {Kieboom}}{{Heise} et~al.}{1985}]{1985A&A...148L..14H}
{Heise} J.,  {Brinkman} A.~C.,  {Gronenschild} E.,  {Watson} M.,  {King} A.~R.,
   {Stella} L.,   {Kieboom} K.,  1985, \aap, \href
  {https://ui.adsabs.harvard.edu/abs/1985A&A...148L..14H} {148, L14}

\bibitem[\protect\citeauthoryear{{Hellier}}{{Hellier}}{2001}]{2001cvs..book.....H}
{Hellier} C.,  2001, {Cataclysmic Variable Stars}

\bibitem[\protect\citeauthoryear{{Hong}}{{Hong}}{2012}]{2012MNRAS.427.1633H}
{Hong} J.,  2012, \mn@doi [\mnras] {10.1111/j.1365-2966.2012.22079.x}, \href
  {https://ui.adsabs.harvard.edu/abs/2012MNRAS.427.1633H} {427, 1633}

\bibitem[\protect\citeauthoryear{{Hong}, {van den Berg}, {Grindlay}  \&
  {Laycock}}{{Hong} et~al.}{2009}]{2009ApJ...706..223H}
{Hong} J.~S.,  {van den Berg} M.,  {Grindlay} J.~E.,   {Laycock} S.,  2009,
  \mn@doi [\apj] {10.1088/0004-637X/706/1/223}, \href
  {https://ui.adsabs.harvard.edu/abs/2009ApJ...706..223H} {706, 223}

\bibitem[\protect\citeauthoryear{{Hong}, {van den Berg}, {Grindlay},
  {Servillat}  \& {Zhao}}{{Hong} et~al.}{2012}]{2012ApJ...746..165H}
{Hong} J.,  {van den Berg} M.,  {Grindlay} J.~E.,  {Servillat} M.,   {Zhao} P.,
   2012, \mn@doi [\apj] {10.1088/0004-637X/746/2/165}, \href
  {https://ui.adsabs.harvard.edu/abs/2012ApJ...746..165H} {746, 165}

\bibitem[\protect\citeauthoryear{{Knigge}, {Baraffe}  \& {Patterson}}{{Knigge}
  et~al.}{2011}]{2011ApJS..194...28K}
{Knigge} C.,  {Baraffe} I.,   {Patterson} J.,  2011, \mn@doi [\apjs]
  {10.1088/0067-0049/194/2/28}, \href
  {https://ui.adsabs.harvard.edu/abs/2011ApJS..194...28K} {194, 28}

\bibitem[\protect\citeauthoryear{{Leahy}, {Darbro}, {Elsner}, {Weisskopf},
  {Sutherland}, {Kahn}  \& {Grindlay}}{{Leahy}
  et~al.}{1983}]{1983ApJ...266..160L}
{Leahy} D.~A.,  {Darbro} W.,  {Elsner} R.~F.,  {Weisskopf} M.~C.,  {Sutherland}
  P.~G.,  {Kahn} S.,   {Grindlay} J.~E.,  1983, \mn@doi [\apj]
  {10.1086/160766}, \href
  {https://ui.adsabs.harvard.edu/abs/1983ApJ...266..160L} {266, 160}

\bibitem[\protect\citeauthoryear{{Lomb}}{{Lomb}}{1976}]{1976Ap&SS..39..447L}
{Lomb} N.~R.,  1976, \mn@doi [\apss] {10.1007/BF00648343}, \href
  {https://ui.adsabs.harvard.edu/abs/1976Ap&SS..39..447L} {39, 447}

\bibitem[\protect\citeauthoryear{{Morihana}, {Tsujimoto}, {Yoshida}  \&
  {Ebisawa}}{{Morihana} et~al.}{2013}]{2013ApJ...766...14M}
{Morihana} K.,  {Tsujimoto} M.,  {Yoshida} T.,   {Ebisawa} K.,  2013, \mn@doi
  [\apj] {10.1088/0004-637X/766/1/14}, \href
  {https://ui.adsabs.harvard.edu/abs/2013ApJ...766...14M} {766, 14}

\bibitem[\protect\citeauthoryear{{Mukai}}{{Mukai}}{2017}]{2017PASP..129f2001M}
{Mukai} K.,  2017, \mn@doi [\pasp] {10.1088/1538-3873/aa6736}, \href
  {https://ui.adsabs.harvard.edu/abs/2017PASP..129f2001M} {129, 062001}

\bibitem[\protect\citeauthoryear{{Muno} et~al.,}{{Muno}
  et~al.}{2003a}]{2003ApJ...589..225M}
{Muno} M.~P.,  et~al., 2003a, \mn@doi [\apj] {10.1086/374639}, \href
  {https://ui.adsabs.harvard.edu/abs/2003ApJ...589..225M} {589, 225}

\bibitem[\protect\citeauthoryear{{Muno}, {Baganoff}, {Bautz}, {Brand t},
  {Garmire}  \& {Ricker}}{{Muno} et~al.}{2003b}]{2003ApJ...599..465M}
{Muno} M.~P.,  {Baganoff} F.~K.,  {Bautz} M.~W.,  {Brand t} W.~N.,  {Garmire}
  G.~P.,   {Ricker} G.~R.,  2003b, \mn@doi [\apj] {10.1086/379244}, \href
  {https://ui.adsabs.harvard.edu/abs/2003ApJ...599..465M} {599, 465}

\bibitem[\protect\citeauthoryear{{Muno} et~al.,}{{Muno}
  et~al.}{2004}]{2004ApJ...613.1179M}
{Muno} M.~P.,  et~al., 2004, \mn@doi [\apj] {10.1086/423164}, \href
  {https://ui.adsabs.harvard.edu/abs/2004ApJ...613.1179M} {613, 1179}

\bibitem[\protect\citeauthoryear{{Muno}, {Bauer}, {Bandyopadhyay}  \&
  {Wang}}{{Muno} et~al.}{2006}]{2006ApJS..165..173M}
{Muno} M.~P.,  {Bauer} F.~E.,  {Bandyopadhyay} R.~M.,   {Wang} Q.~D.,  2006,
  \mn@doi [\apjs] {10.1086/504798}, \href
  {https://ui.adsabs.harvard.edu/abs/2006ApJS..165..173M} {165, 173}

\bibitem[\protect\citeauthoryear{{Muno} et~al.,}{{Muno}
  et~al.}{2009}]{2009ApJS..181..110M}
{Muno} M.~P.,  et~al., 2009, \mn@doi [\apjs] {10.1088/0067-0049/181/1/110},
  \href {https://ui.adsabs.harvard.edu/abs/2009ApJS..181..110M} {181, 110}

\bibitem[\protect\citeauthoryear{{Norton}, {Beardmore}  \& {Taylor}}{{Norton}
  et~al.}{1996}]{1996MNRAS.280..937N}
{Norton} A.~J.,  {Beardmore} A.~P.,   {Taylor} P.,  1996, \mn@doi [\mnras]
  {10.1093/mnras/280.3.937}, \href
  {https://ui.adsabs.harvard.edu/abs/1996MNRAS.280..937N} {280, 937}

\bibitem[\protect\citeauthoryear{{Nucita}, {Kuulkers}, {Maiolo}, {de Paolis},
  {Ingrosso}  \& {Vetrugno}}{{Nucita} et~al.}{2011}]{2011A&A...536A..75N}
{Nucita} A.~A.,  {Kuulkers} E.,  {Maiolo} B.~M.~T.,  {de Paolis} F.,
  {Ingrosso} G.,   {Vetrugno} D.,  2011, \mn@doi [\aap]
  {10.1051/0004-6361/201117572}, \href
  {https://ui.adsabs.harvard.edu/abs/2011A&A...536A..75N} {536, A75}

\bibitem[\protect\citeauthoryear{{Pala} et~al.,}{{Pala}
  et~al.}{2020}]{2020MNRAS.494.3799P}
{Pala} A.~F.,  et~al., 2020, \mn@doi [\mnras] {10.1093/mnras/staa764}, \href
  {https://ui.adsabs.harvard.edu/abs/2020MNRAS.494.3799P} {494, 3799}

\bibitem[\protect\citeauthoryear{{Perez} et~al.,}{{Perez}
  et~al.}{2015}]{2015Natur.520..646P}
{Perez} K.,  et~al., 2015, \mn@doi [\nat] {10.1038/nature14353}, \href
  {https://ui.adsabs.harvard.edu/abs/2015Natur.520..646P} {520, 646}

\bibitem[\protect\citeauthoryear{{Pretorius}, {Knigge}  \&
  {Schwope}}{{Pretorius} et~al.}{2013}]{2013MNRAS.432..570P}
{Pretorius} M.~L.,  {Knigge} C.,   {Schwope} A.~D.,  2013, \mn@doi [\mnras]
  {10.1093/mnras/stt499}, \href
  {https://ui.adsabs.harvard.edu/abs/2013MNRAS.432..570P} {432, 570}

\bibitem[\protect\citeauthoryear{{Rappaport}, {Cash}, {Doxsey}, {McClintock}
  \& {Moore}}{{Rappaport} et~al.}{1974}]{1974ApJ...187L...5R}
{Rappaport} S.,  {Cash} W.,  {Doxsey} R.,  {McClintock} J.,   {Moore} G.,
  1974, \mn@doi [\apjl] {10.1086/181378}, \href
  {https://ui.adsabs.harvard.edu/abs/1974ApJ...187L...5R} {187, L5}

\bibitem[\protect\citeauthoryear{{Reid}, {Menten}, {Zheng}, {Brunthaler}  \&
  {Xu}}{{Reid} et~al.}{2009}]{2009ApJ...705.1548R}
{Reid} M.~J.,  {Menten} K.~M.,  {Zheng} X.~W.,  {Brunthaler} A.,   {Xu} Y.,
  2009, \mn@doi [\apj] {10.1088/0004-637X/705/2/1548}, \href
  {https://ui.adsabs.harvard.edu/abs/2009ApJ...705.1548R} {705, 1548}

\bibitem[\protect\citeauthoryear{{Revnivtsev}, {Sazonov}, {Churazov}, {Forman},
  {Vikhlinin}  \& {Sunyaev}}{{Revnivtsev} et~al.}{2009}]{2009Natur.458.1142R}
{Revnivtsev} M.,  {Sazonov} S.,  {Churazov} E.,  {Forman} W.,  {Vikhlinin} A.,
   {Sunyaev} R.,  2009, \mn@doi [\nat] {10.1038/nature07946}, \href
  {https://ui.adsabs.harvard.edu/abs/2009Natur.458.1142R} {458, 1142}

\bibitem[\protect\citeauthoryear{{Revnivtsev}, {Sazonov}, {Forman}, {Churazov}
  \& {Sunyaev}}{{Revnivtsev} et~al.}{2011}]{2011MNRAS.414..495R}
{Revnivtsev} M.,  {Sazonov} S.,  {Forman} W.,  {Churazov} E.,   {Sunyaev} R.,
  2011, \mn@doi [\mnras] {10.1111/j.1365-2966.2011.18411.x}, \href
  {https://ui.adsabs.harvard.edu/abs/2011MNRAS.414..495R} {414, 495}

\bibitem[\protect\citeauthoryear{{Ritter} \& {Kolb}}{{Ritter} \&
  {Kolb}}{2003}]{2003A&A...404..301R}
{Ritter} H.,  {Kolb} U.,  2003, \mn@doi [\aap] {10.1051/0004-6361:20030330},
  \href {https://ui.adsabs.harvard.edu/abs/2003A&A...404..301R} {404, 301}

\bibitem[\protect\citeauthoryear{{Sazonov}, {Revnivtsev}, {Gilfanov},
  {Churazov}  \& {Sunyaev}}{{Sazonov} et~al.}{2006}]{2006A&A...450..117S}
{Sazonov} S.,  {Revnivtsev} M.,  {Gilfanov} M.,  {Churazov} E.,   {Sunyaev} R.,
   2006, \mn@doi [\aap] {10.1051/0004-6361:20054297}, \href
  {https://ui.adsabs.harvard.edu/abs/2006A&A...450..117S} {450, 117}

\bibitem[\protect\citeauthoryear{{Scargle}}{{Scargle}}{1982}]{1982ApJ...263..835S}
{Scargle} J.~D.,  1982, \mn@doi [\apj] {10.1086/160554}, \href
  {https://ui.adsabs.harvard.edu/abs/1982ApJ...263..835S} {263, 835}

\bibitem[\protect\citeauthoryear{{Schuster}}{{Schuster}}{1898}]{1898TeMag...3...13S}
{Schuster} A.,  1898, \mn@doi [Terrestrial Magnetism (Journal of Geophysical
  Research)] {10.1029/TM003i001p00013}, \href
  {https://ui.adsabs.harvard.edu/abs/1898TeMag...3...13S} {3, 13}

\bibitem[\protect\citeauthoryear{{Schwarzenberg-Czerny}}{{Schwarzenberg-Czerny}}{1996}]{1996ApJ...460L.107S}
{Schwarzenberg-Czerny} A.,  1996, \mn@doi [\apjl] {10.1086/309985}, \href
  {https://ui.adsabs.harvard.edu/abs/1996ApJ...460L.107S} {460, L107}

\bibitem[\protect\citeauthoryear{{Stellingwerf}}{{Stellingwerf}}{1978}]{1978ApJ...224..953S}
{Stellingwerf} R.~F.,  1978, \mn@doi [\apj] {10.1086/156444}, \href
  {https://ui.adsabs.harvard.edu/abs/1978ApJ...224..953S} {224, 953}

\bibitem[\protect\citeauthoryear{{Wang}, {Gotthelf}  \& {Lang}}{{Wang}
  et~al.}{2002}]{2002Natur.415..148W}
{Wang} Q.~D.,  {Gotthelf} E.~V.,   {Lang} C.~C.,  2002, \mn@doi [\nat]
  {10.1038/415148a}, \href
  {https://ui.adsabs.harvard.edu/abs/2002Natur.415..148W} {415, 148}

\bibitem[\protect\citeauthoryear{{Wevers} et~al.,}{{Wevers}
  et~al.}{2016}]{2016MNRAS.462L.106W}
{Wevers} T.,  et~al., 2016, \mn@doi [\mnras] {10.1093/mnrasl/slw141}, \href
  {https://ui.adsabs.harvard.edu/abs/2016MNRAS.462L.106W} {462, L106}

\bibitem[\protect\citeauthoryear{{Xu}, {Wang}  \& {Li}}{{Xu}
  et~al.}{2016}]{2016ApJ...818..136X}
{Xu} X.-J.,  {Wang} Q.~D.,   {Li} X.-D.,  2016, \mn@doi [\apj]
  {10.3847/0004-637X/818/2/136}, \href
  {https://ui.adsabs.harvard.edu/abs/2016ApJ...818..136X} {818, 136}

\bibitem[\protect\citeauthoryear{{Xu}, {Li}, {Zhu}, {Cheng}, {Li}  \&
  {Yu}}{{Xu} et~al.}{2019}]{2019ApJ...882..164X}
{Xu} X.-J.,  {Li} Z.,  {Zhu} Z.,  {Cheng} Z.,  {Li} X.-d.,   {Yu} Z.-l.,  2019,
  \mn@doi [\apj] {10.3847/1538-4357/ab32df}, \href
  {https://ui.adsabs.harvard.edu/abs/2019ApJ...882..164X} {882, 164}

\bibitem[\protect\citeauthoryear{{Zhu}, {Li}  \& {Morris}}{{Zhu}
  et~al.}{2018}]{2018ApJS..235...26Z}
{Zhu} Z.,  {Li} Z.,   {Morris} M.~R.,  2018, \mn@doi [\apjs]
  {10.3847/1538-4365/aab14f}, \href
  {https://ui.adsabs.harvard.edu/abs/2018ApJS..235...26Z} {235, 26}

\bibitem[\protect\citeauthoryear{{van den Berg}, {Hong}  \& {Grindlay}}{{van
  den Berg} et~al.}{2009}]{2009ApJ...700.1702V}
{van den Berg} M.,  {Hong} J.~S.,   {Grindlay} J.~E.,  2009, \mn@doi [\apj]
  {10.1088/0004-637X/700/2/1702}, \href
  {https://ui.adsabs.harvard.edu/abs/2009ApJ...700.1702V} {700, 1702}

\makeatother
\end{thebibliography}
\bibliographystyle{mnras}
\appendix
\section{A brief introduction to the Gregory-Loredo algorithm}\label{GL}
The basic rules for Bayesian probabilities are the sum rule,
\begin{equation}
p(H_i|I)+p(\bar{H_i}|I)	=1,
\end{equation}
and the product rule,
\begin{equation}\label{2.2}
p(H_i,D|I)=p(H_i|I)\cdot p(D|H_i,I)=p(D|I)\cdot p(H_i|D,I).
\end{equation}
From Eqn.~\ref{2.2} we can derive Bayes's theorem,
\begin{equation}\label{2.5} 
p(H_i|D,I)=p(H_i|I)\cdot {p(D|H_i,I)\over p(D|I)}.
\end{equation}
The symbols here follow \cite{1992ApJ...398..146G}.
Specifically, $p$ is the Bayesian posterior probability, $H_i$ denotes the $i$-th hypothesis, $D$ for the data, and $I$ for the ensemble of all hypotheses, i.e., all the models used.
The GL algorithm employs a stepwise function to detect periodic signal. Each model has $(m+2)$ parameters: the angular frequency $\omega={2\pi/P}$ ($P$ is the period), the phase parameter $\phi$, and $m$ values of $r_k$, which denotes the count rate in each phase bin where $k$=1 to $m$. In the following we replace $H_i$ by $M_i$ to denote the model where $i$ represents the number of bins in the stepwise model. Then the Bayes's theorem can be written as,
 \begin{equation}\label{2.11}
 p(M_i|D,I)=p(M_i|I)\cdot {p(D|M_i,I)\over p(D|I)}.
 \end{equation} 
We can write $I=M_1+M_2+M_3+\cdots$, where ``+'' stands for ``or''. Thus the proposition ($M_i$, $I$) is true if and only if model $M_i$ is true, i.e., ($M_i$, $I$) = $M_i$. The GL algorithm defines an odds ratio for model comparison,
\begin{equation}\label{2.12}
O_{ij}={p(M_i|D,I)\over p(M_j|D,I)}={p(M_i|I)\over p(M_j|I)}\cdot {p(D|M_i)\over p(D|M_j)}.
\end{equation}
Note that $M_1$ means a constant model, while $M_i$ ($i$ = 2, 3, 4...$N_{\rm mod}$, where $N_{\rm mod}$ is the total number of models considered) represents a periodic model. The probability for each model can be deduced from Eqn.~\ref{2.12},
\begin{equation}\label{6}
p(M_i|D,I)=O_{i1}\cdot p(M_1|D,I),
\end{equation}
thus
\begin{equation}\label{7}
p(M_1|D,I)={{\sum_{j=1}^{N_{\rm mod}} p(M_j|D,I)}\over {\sum_{j=1}^{N_{\rm mod}} O_{j1}}}
={1\over {\sum_{j=1}^{N_{\rm mod}} O_{j1}}}.
\end{equation}
Substituting Eqn.~\ref{7} into Eqn.~\ref{6}, we have
\begin{equation}\label{2.13}
p(M_i|D,I)={O_{i1}\over {\sum_{j=1}^{N_{\rm mod}} O_{j1}}}.
\end{equation}
Then the probability of a periodic signal is
\begin{equation}\label{5.2}
p(M_m(m>1)|D,I)={{\sum_{m=2}^{m_{\rm max}} O_{m1}}\over {1+\sum_{m=2}^{m_{\rm max}} O_{m1}}},
\end{equation}
where $m_{\rm max}$ is the maximum value of $m$.
The odds ratio can be calculated from the probability of the model,
\begin{equation}
O_{m1}={{p(M_m|D,I)}\over {p(M_1|D,I)}}.
\end{equation}
Using Bayes's theorem (Eqn.~\ref{2.11}), 
\begin{equation}\label{11}
O_{m1}={{p(M_m|I)\cdot p(D|M_m)}\over {p(M_1|I)\cdot p(D|M_1)}}.
\end{equation}
Following the assignment by \citet{1992ApJ...398..146G}, we assume that the periodic and aperiodic signals have the same probability. Then the priors for the models can be written explicitly as,
\begin{equation}\label{12}
p(M_1|I)={1\over 2},	
\end{equation}
\begin{equation}\label{13}
p(M_m|I)={1\over {2\nu}}, \quad \nu =m_{\rm max}-1.
\end{equation}

For astronomical data of Poisson distribution, it can be shown that (Equation 5.27 in \citealp{1992ApJ...398..146G}),
\begin{equation}\label{5.27}
\begin{split}
p(D|M_m)={{{\Delta t}^N (m-1)! N! \gamma(N+1,A_{\rm max})T}\over{2\pi A_{\rm max} (N+m-1)!T^{N+1} \ln(\omega_{\rm hi}/\omega_{\rm lo})}}\\
\times {\int_{\omega_{\rm lo}}^{\omega_{\rm hi}}}{{d\omega}\over{\omega}}\times {\int_0^{2\pi}d\phi{m^N\over W_m(\omega,\phi)}},	
\end{split}
\end{equation}
where $\omega_{\rm lo}$ and $\omega_{\rm hi}$ are the lower and upper bounds of the frequency range, and $T$ is the total duration of observing intervals.
It should be emphasized that the above equation holds for the case in which the period and phase are both unknown.
Substituting $m=1$ into Eqn.~\ref{5.27}, we have
\begin{equation}\label{15}
p(D|M_1)={{{\Delta t}^N N!\gamma(N+1,A_{\rm max})T}\over {A_{\rm max}N!T^{N+1}}}.
\end{equation}
Substituting Eqns.~\ref{12}, \ref{13}, \ref{5.27} and \ref{15} into Eqn.~\ref{11}, the odds ratio can be written as follows, which is the same as Equation 5.28 in \cite{1992ApJ...398..146G},
\begin{equation}\label{A16}
\begin{split}
O_{m1}={1\over{2\pi \nu \ln(\omega_{\rm hi}/\omega_{\rm lo})}} {{N+m-1}	\choose N}^{-1}\times {\int_{\omega_{\rm lo}}^{\omega_{\rm hi}}}{{d\omega}\over{\omega}}\\
\times {\int_0^{2\pi}d\phi{m^N\over W_m(\omega,\phi)}} 
\end{split}
\end{equation}

Astronomical data are often subject to observational gaps. 
This may result in unevenly covered phase bins, leading to spurious detections especially at low frequencies. \cite{1992ApJ...398..146G} provides a solution to this problem, by introducing a weighting factor 
\begin{equation}\label{A17}
S(\omega,\phi)={\prod_{j=1}^m s_j^{-n_j}},
\end{equation}
\begin{equation}
s_j(\omega,\phi)={{\tau_{j}(\omega,\phi)}\over {T/m}},
\end{equation}
where ${\tau_{j}(\omega,\phi)}$ denotes the exposure time in each phase bin.
Then the odds ratio should be modified as,
\begin{equation}\label{A19}
\begin{split}
O_{m1}={1\over{2\pi \nu \ln(\omega_{\rm hi}/\omega_{\rm lo})}} {{N+m-1}	\choose N}^{-1}\times {\int_{\omega_{\rm lo}}^{\omega_{\rm hi}}}{{d\omega}\over{\omega}}\\
\times {\int_0^{2\pi}d\phi{S(\omega,\phi)m^N\over W_m(\omega,\phi)}}.
\end{split} 
\end{equation}
Ultimately, the probability of a periodic signal is,
\begin{equation}\label{A20}
p({\rm periodic})={{\sum_{m=2}^{m_{\rm max}} O_{m1}}\over {1+\sum_{m=2}^{m_{\rm max}} O_{m1}}}.
\end{equation}
The posterior probability as a function of the frequency,
\begin{equation}\label{A21}
O_{m1}(\omega)={1\over{2\pi \nu}} {{N+m-1}	\choose N}^{-1} \times {\int_0^{2\pi}d\phi{S(\omega,\phi)m^N\over W_m(\omega,\phi)}},
\end{equation}
can be solved to find the period $P=2\pi /\omega$, when $O_{m1}(\omega)$ takes the maximum value.

\section{Additional figures of the periodic X-ray sources}\label{appen:fig}
  \begin{figure*}
    \centering
    \includegraphics[page=1,scale=0.90,trim=0 100 0 20,clip]{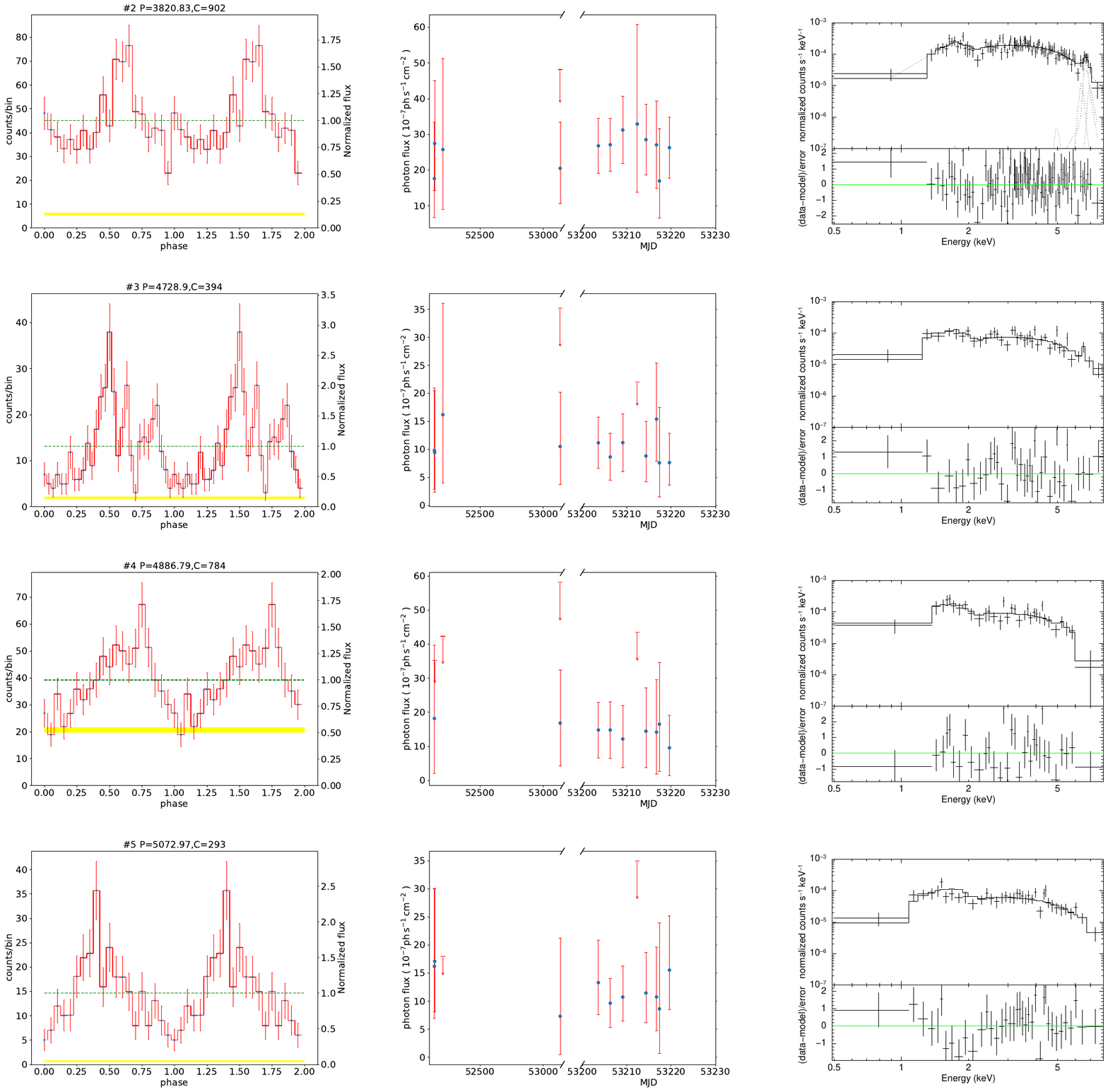}
    \caption{Each row shows one periodic source. {\it Left}: The 1--8 keV phase-folded light curve at the modulation period.
The green dashed line represents the mean count rate, whereas the yellow strip represents the local background, the width of which represents 1\,$\sigma$ Poisson error.
{\it Middle}: the 1--8 keV long-term, inter-observation light curve. Arrows represent 3\,$\sigma$ upper limits. 
{\it Right}: Source spectrum and the best-fit model. 
    \label{fig:Figure_p}}
  \end{figure*}
  
  \begin{figure*}
    \centering
    \includegraphics[page=2,scale=0.90,trim=0 100 0 20,clip]{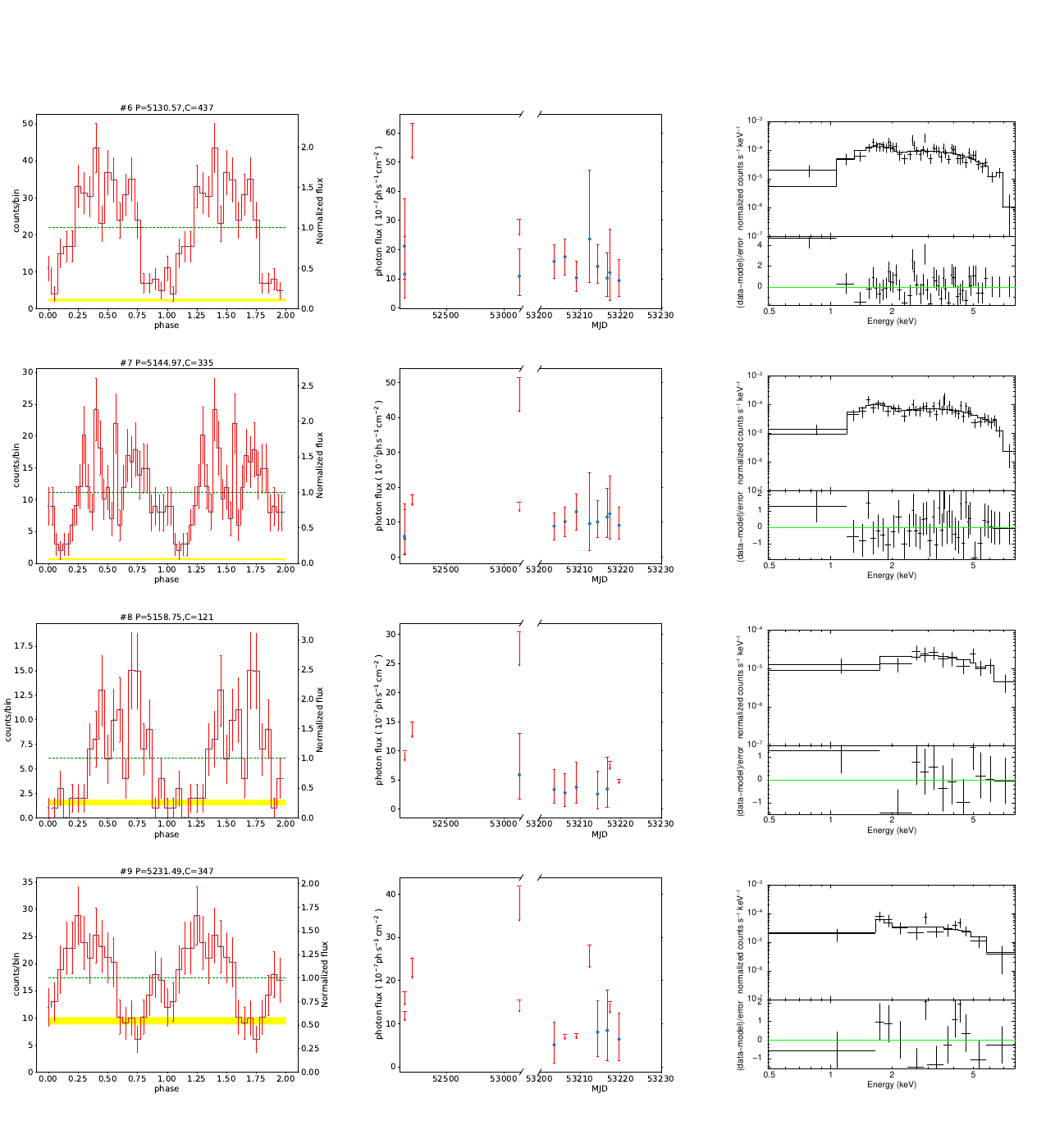}
    \caption{Continued}
  \end{figure*}

  \begin{figure*}
    \centering
    \includegraphics[page=3,scale=0.90,trim=0 100 0 20,clip]{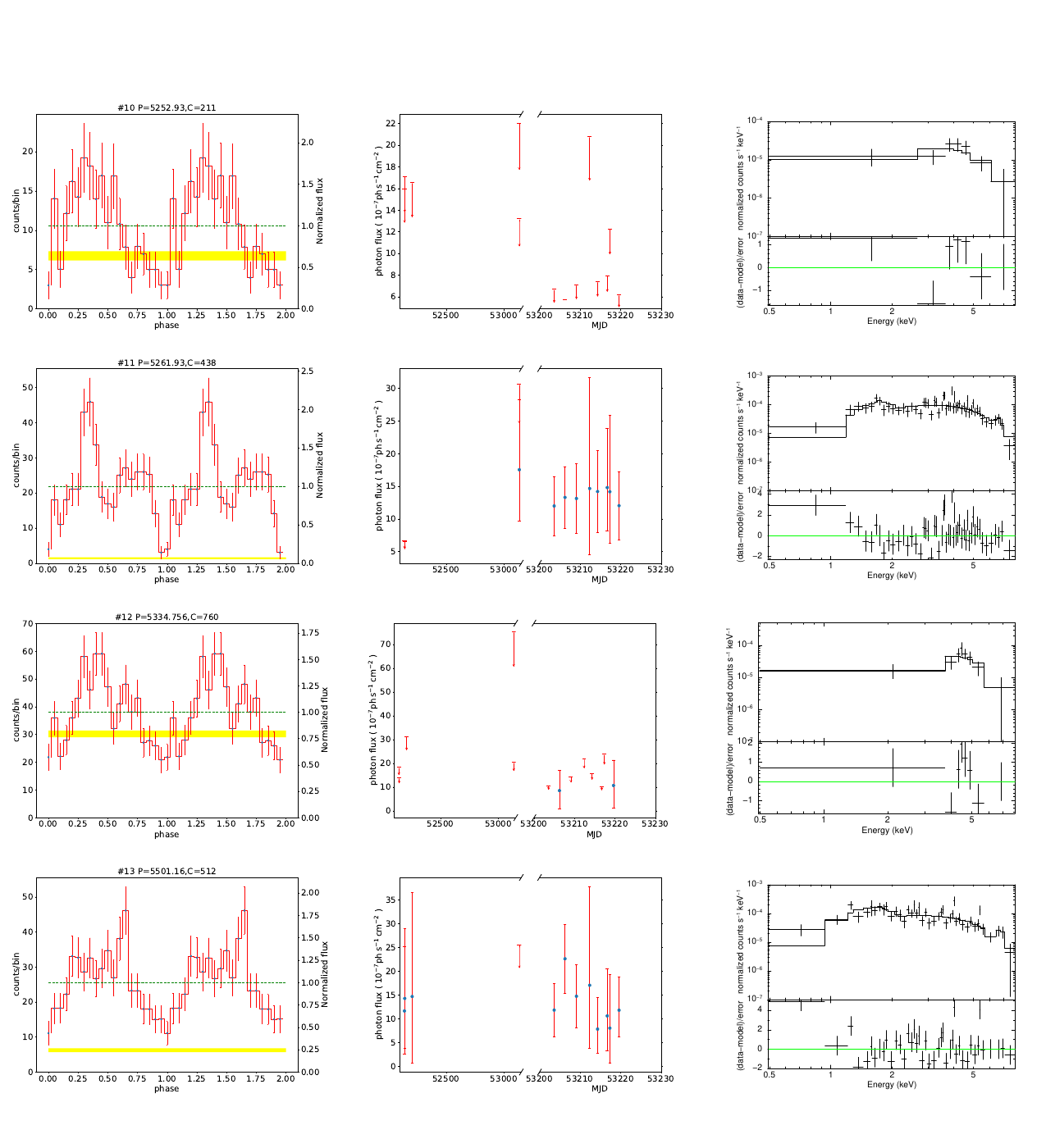}
    \caption{Continued}
  \end{figure*}
  
   \begin{figure*}
    \centering
    \includegraphics[page=4,scale=0.90,trim=0 100 0 20,clip]{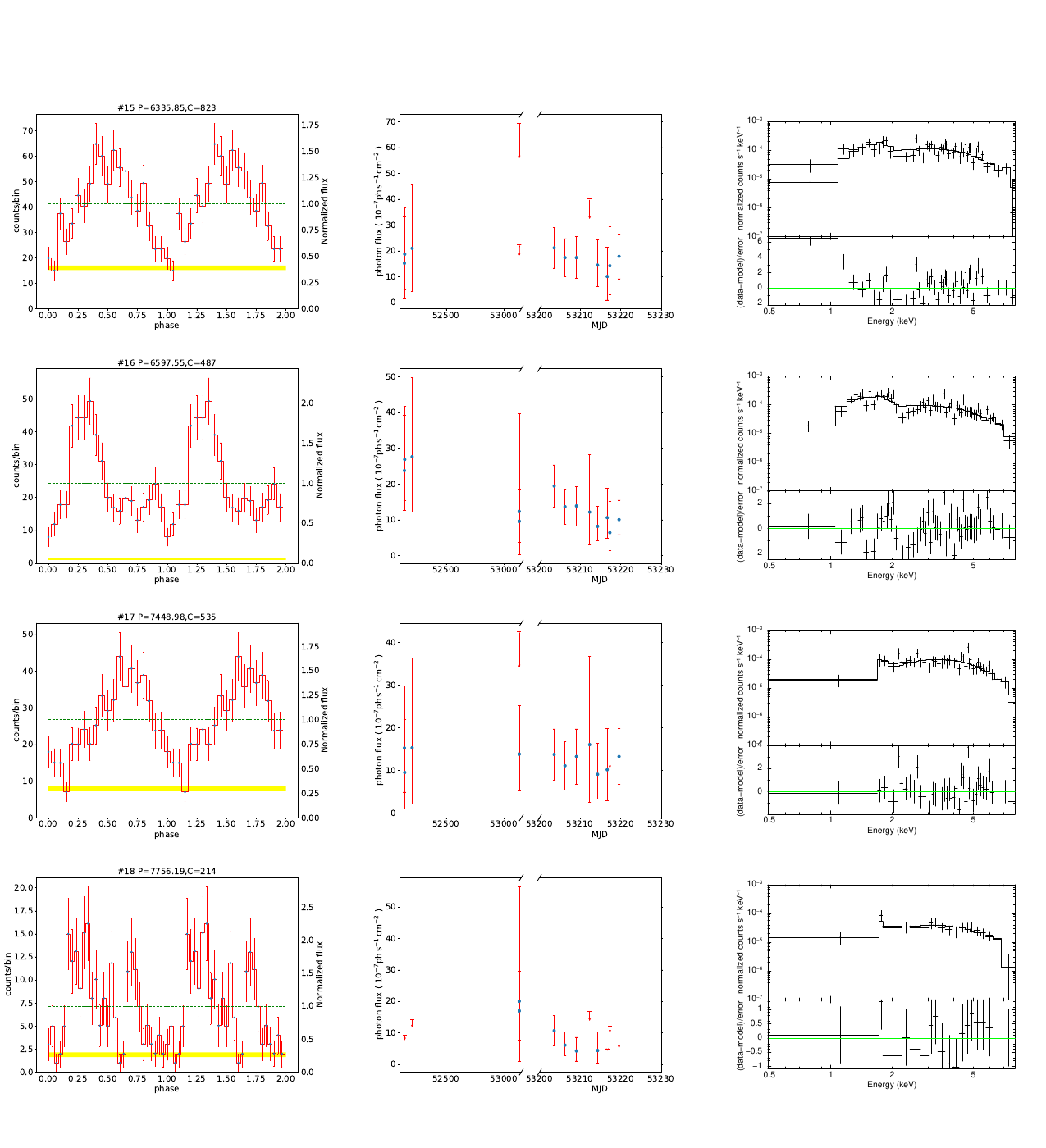}
    \caption{Continued}
  \end{figure*}
  
  \begin{figure*}
    \centering
    \includegraphics[page=5,scale=0.90,trim=0 100 0 20,clip]{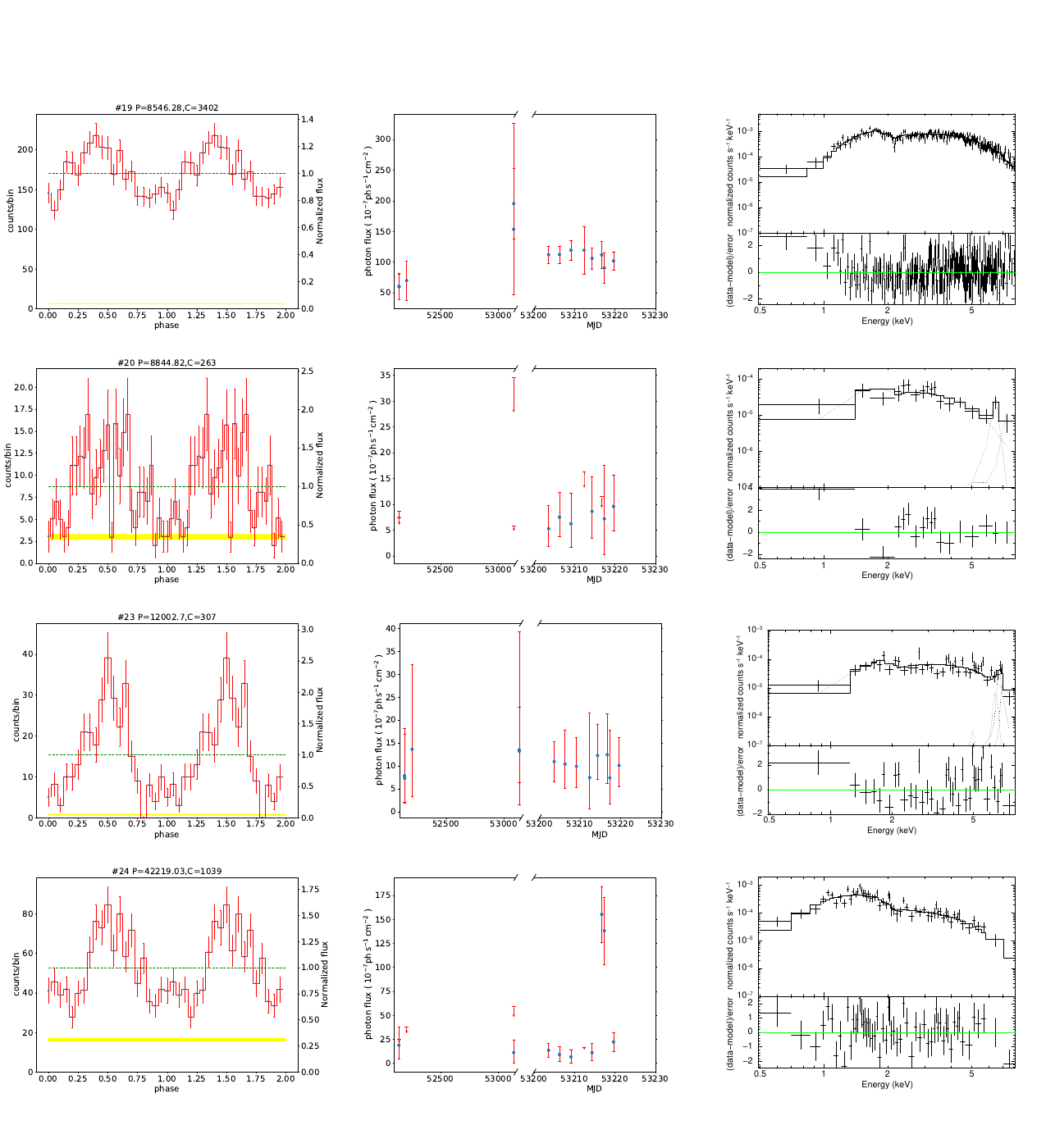}
    \caption{Continued}
  \end{figure*}
  
  \begin{figure*}
    \centering
    \includegraphics[page=6,scale=0.90,trim=0 100 0 20,clip]{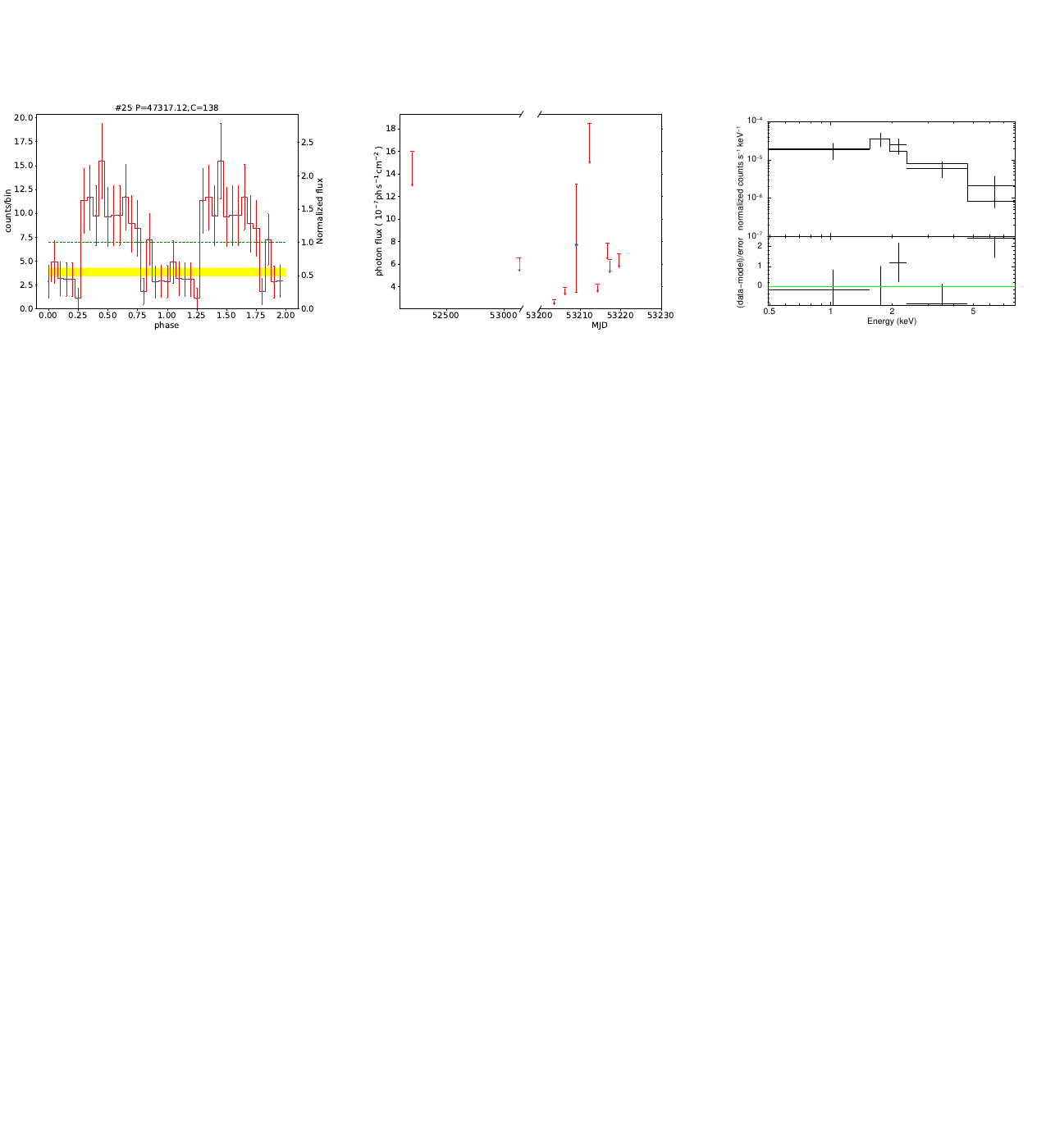}
    \caption{Continued}
  \end{figure*}
  


\end{document}